# A conformal gauge theory of solids: insights into a class of electromechanical and magnetomechanical phenomena

Pranesh Roy[1], J N Reddy[3,*] and Debasish Roy[1,2]

[1]*Computational Mechanics Lab, Department of Civil Engineering*

[2]*Centre of Excellence in Advanced Mechanics of Materials*

*Indian Institute of Science, Bangalore 560012, India*

[3]*Advanced Computational Mechanics Lab, Department of Mechanical Engineering*

*Texas A&M University, College Station, Texas 77843-3123*

(*Corresponding author; email: jnreddy@tamu.edu*)

**Abstract**

A gauge theory of solids with conformal symmetry is formulated to model various electromechanical and magnetomechanical coupling phenomena. If the pulled back metric of the current configuration (the right Cauchy-Green tensor) is scaled with a constant, the volumetric part of the Lagrange density changes while the isochoric part remains invariant. However, upon a position dependent scaling, the isochoric part also loses invariance. In order to restore the invariance of the isochoric part, a 1-form compensating field is introduced and the notion of a gauge covariant derivative is utilized to minimally replace the Lagrangian. In view of obvious similarities with the Weyl geometry, the Weyl condition is imposed through the Lagrangian and a minimal coupling is employed so the 1-form could evolve. On deriving the Euler-Lagrange equations based on Hamilton's principle, we observe a close similarity with the governing equations for flexoelectricity under isochoric deformation if the exact part of 1-form is interpreted as the electric field and the anti-exact part as the polarization vector. Next, we model piezoelectricity and electrostriction phenomena by contracting the Weyl condition in various ways. Applying the Hodge decomposition theorem on the 1-form which leads to the curl of a pseudo-vector field and a vector field, we also model magnetomechanical phenomena. Identifying the pseudo-vector field with magnetic potential and the vector part with magnetization, flexomagnetism, piezomagnetism and magnetostriction phenomena under isochoric deformation are also modeled. Finally, we consider an analytical solution of the equations for piezoelectricity to provide an illustration on the insightful information that the present approach potentially provides.

1. Introduction

Electromechanical and magnetomechanical coupling phenomena continue to receive significant research attention due to their large-scale industrial applications. Piezoelectricity is one such electromechanical phenomenon that has been extensively investigated and exploited for energy harvesting purposes. Many naturally occurring materials such as quartz, ceramics (e.g. lead zirconate titanate (PZT)), polymers (Polyvinylidene fluoride (PVDF)) exhibit piezoelectric behaviour and are exploited in sensors and actuators, medical instruments and various energy harvesting devices. Another electromechanical phenomenon, flexoelectricity, has relevance in the nanoscale, where strain gradients being typically very high are suggestive of a remarkably enhanced harvesting power. Most available theoretical models for electromechanical and magnetomechanical phenomena start with an energy functional or Hamiltonian with coupling terms, which are empirically introduced without a basis in the deep underlying symmetry requirements of the Lagrangian. Such a heuristic modelling principle lacks in its predictive quality. For instance, electromechanical and magnetomechanical response under large deformation remains an open challenge. In order to better understand these phenomena observed in solid materials from a more fundamental perspective with a view to deriving a more rational and unified predictive model, one may use the gauge theory of solids wherein certain local symmetries are imposed on the Lagrangian.

Lagoudas and Edelen (1989) have dwelt on a gauge theoretic approach to explain the dynamic behavior of solids with (a class of) continuously distributed defects. As we know, the classical Lagrange density of elasticity theory is invariant under the global (homogeneous) action of certain spatial and material symmetry groups, e.g. rotation and translation. The action of the spatial symmetry group $G_s = SO(3) \rhd T(3)$ is given by transforming the deformation map ($\chi$) at a given time $t$ in the following way: $`\chi = Q\chi + b$ and $`t = t$ with $Q \in SO(3)$ and $b \in T(3)$. It describes rigid body motion of the deformed configuration and leaves the stored elastic energy of the material body unaffected. Lagoudas and Edelen (1989) have also looked into this transformation from another point of view: a change of position of the observer at time $t$, whilst keeping the deformed configuration the same. In other words, the spatial coordinate cover is transformed as: $`x = Qx + b$ and $`t = t$. According to the principle of frame indifference, the Lagrange density remains invariant under this transformation. On the other hand, the action of the material symmetry group $G_m = \{SO(3) \triangleright T(3)\} \times T(1)$ is prescribed through a transformation of reference configurations as: $`X^A = R^A_B X^B + T^A$ and $`X^4 = X^4 + T^4$, where, $A, B \in \{1, 2, 3\}$, $R \in SO(3)$ and $T \in T(1)$, under which the reference Lagrange density is invariant.

However, if the parameters of the spatial symmetry group ($G_s$) and material symmetry group ($G_m$) are made inhomogeneous or local, i.e. they are made to depend on position and time, then the invariance of the classical Lagrange density no longer holds. In order to restore the invariance under the local action of $G_s$, following the Yang-Mills procedure, Kadić and Edelen (1982) modified the definition of partial derivatives to covariant ones, where a nontrivial gauge connection $\Gamma$ was introduced and a transformation of $\Gamma$ under the action of $G_s$ derived. This connection may also be viewed as the Yang-Mills compensating potential which arises through a breaking of the homogeneity of the group action. The operation of modifying the definition of partial derivative is called minimal replacement. Application of this minimal replacement construct in the elasticity theory leads to a modified definition of deformation gradient called distortion tensor and the material velocity is modified to distortional velocity. If the new Lagrange density is constructed through the modified deformation gradient, then it remains invariant under the position and time dependent transformations. Two viewpoints exist in this context, viz. principles of generalized frame indifference and generalized objectivity. While generalized frame indifference deals with minimal replacement of the derivative to compensate for local rigid body motion of the observer and constructs Lagrange density with the modified deformation gradient, generalized objectivity applies minimal replacement to compensate for position dependent transformation of the deformed configuration and then classical objectivity principle. Similarly, in order to restore the invariance of the Lagrange density under local action of $G_m$, Lagoudas and Edelen (1989) proposed a minimal replacement construct via the introduction of a nontrivial affine connection, which apart from modifying the definition of partial derivatives through covariant derivatives, modifies the four-dimensional infinitesimal volume element in the reference configuration.

Once the kinematics are defined through minimal replacement, the next step is to prescribe the evolution equations of the coefficients of the nontrivial affine connection. However, as these new field variables appear algebraically in the Lagrangian, upon application of Hamilton's principle i.e. the stationarity of the action with respect to the new field variables, unrealistic solutions arise. In order to overcome this difficulty, following Yang and Mills, Kadić and Edelen (1982) introduced additional gauge invariant terms in the Lagrangian, which consisted of quadratic forms of the first order derivatives of the compensating potentials. More specifically, these terms contained components of the curvature associated with the connection 1-forms of the gauge groups $G_s$ and $G_m$. The new gauge invariant terms are added to the minimally replaced Lagrangian. This procedure is known as minimal coupling. At this stage, one may find the Euler-Lagrange equations in the form of a set of coupled nonlinear PDEs, which can be solved uniquely with specified boundary conditions. The compensating field variables that appear through breaking of homogeneity under the action of $T(3)$ are interpreted as being associated with dislocations (Kadić and Edelen, 1982). The ones corresponding to homogeneity breaking under the action of $SO(3)$ are thought to represent disclination. Classical elasticity is retrieved in the lowest order approximation and dislocations and disclinations appear in the next consecutive orders respectively. Some important observations from the analysis by Kadić and Edelen (1982) are: the driving force for dislocations is stress, whereas disclinations are driven by both distortions and dislocations, which imply that the creation of disclination requires higher energy as compared to dislocations.

We, in this article, deviate from the class of local symmetries considered above and rather focus on the important role conformal symmetry plays in solid mechanics. Conventionally, for a hyperelastic material, the Lagrange density depends on the invariants of the right Cauchy-Green tensor $\mathbf{C}$. Here, the Lagrange density is typically composed of two parts: one associated with the change in volume and the other isochoric. Also, material response in some cases depends on the gradient of strain. To account for this, one may add another term in the Lagrange density as a function of the covariant derivative of $\mathbf{C}$, which without a loss of generality may be considered isochoric. If one scales the metric, i.e. the right Cauchy-Green tensor $\mathbf{C}$, by a constant factor at all material points (global dilatation), then the volumetric part of the Lagrange density changes whereas the isochoric part remains unaffected. However, if we localize this transformation, i.e. the scaling factor depends on position, then, in addition to the volumetric part, the isochoric part is also affected. In order to formulate a gauge theory for conformal symmetry, we need to modify the definition of the covariant derivative to a gauge covariant one, where we introduce a new 1-form field $\boldsymbol{\gamma}$ and a specific transformation of that field under scale transformation of the metric.

It is worthwhile to put in a remark on the close relation of this gauge construct with Weyl geometry, which is a generalization of Riemann geometry. Riemann geometry is based on a definition of the distance between two infinitesimally separated points as captured by the metric tensor $\mathbf{g}$ and a parallel transport of vectors from one point in the manifold to another along a curve. Here the length and the relative angle between vectors do not change under parallel transport. Weyl geometry differs from Riemann geometry in that, upon parallel transport, the length of a vector changes by a proportionality factor. A consequence of this hypothesis is that, the covariant derivative of $\mathbf{g}$ is not zero as in the Riemannian case, rather it is given in terms of a 1-form field $\boldsymbol{\gamma}$ and the metric $\mathbf{g}$ via Weyl's condition: $\nabla \mathbf{g} = \boldsymbol{\gamma} \otimes \mathbf{g}$. It may be shown that Weyl's condition remains invariant under the transformations $\grave{\mathbf{g}} = e^f \mathbf{g}$ and $\boldsymbol{\gamma}' = \boldsymbol{\gamma} + df$ where $f$ is a smooth scalar valued function.

In the context of gauge theory, we deal with a reference configuration in the Euclidean space and the field variables (e.g. the displacement field) are defined on the reference domain. These field variables are functions of coordinates assigned to the reference configuration. If one wishes to measure the distance between two infinitesimally separated points in the current configuration, then the following equation can be used: $(ds)^2 = \mathbf{C}d\mathbf{X} \cdot d\mathbf{X}$ with $\mathbf{C} = \mathbf{F}^T \mathbf{g} \mathbf{F}$ the right Cauchy Green tensor and $\mathbf{g}$ the metric in the current configuration. Therefore, one may imagine $\mathbf{C}$ as a new metric defined on the reference configuration with respect to which infinitesimal distances in the current configuration can be evaluated at any time $t$. With this viewpoint, the geometry described by the metric $\mathbf{C}$ is not Euclidean, but Riemannian. In the conformal gauge theory of solids, we use the transformation of the metric $\mathbf{C}$ and the additional 1-form field $\boldsymbol{\gamma}$ in the same way as given in Weyl's condition. We minimally replace the covariant derivative as $\bar{\nabla} = \nabla - \boldsymbol{\gamma}$ and use this to construct the gauge invariant quantities in the Lagrangian. As will be shown, the minimally replaced term is written through a contraction of the quantity $\nabla \mathbf{C} - \boldsymbol{\gamma}\mathbf{C}$. Therefore, one may imagine that Weyl's condition is weakly imposed through the minimally replaced Lagrangian. For the minimal coupling operation, we add terms associated with the exterior derivative of $\boldsymbol{\gamma}$ which are invariant under the transformation $\mathbf{C}' = e^f \mathbf{C}$. We note at this point that, under $\mathbf{C}' = e^f \mathbf{C}$, the volumetric (non-isochoric) part changes and the isochoric part remains invariant in the final Lagrangian. One may attempt to write a gauge theory for the volumetric part as well; but we have kept this aspect beyond our current scope. Therefore, we fix the gauge parameter $f$ as $f = 0$ in the volumetric part. Another point of interest here is that $\boldsymbol{\gamma}$ can be decomposed into an exact and an anti-exact part, i.e. $\boldsymbol{\gamma} = d\lambda + \bar{\boldsymbol{\gamma}}$, where $\lambda$ is a zero-form (a real-valued smooth function) and $\bar{\boldsymbol{\gamma}}$ a 1-form. Using Hamilton's principle, one may obtain the Euler-Lagrange equations which in this case are constituted of three coupled equations for $\mathbf{u}$ (the displacement field), $\lambda$ and $\boldsymbol{\gamma}$. These may be solved uniquely after specifying appropriate boundary conditions.

We have thus pointed to a general procedure for formulating a conformal gauge theory that contains two new field variables $\lambda$ and $\boldsymbol{\gamma}$ apart from $\mathbf{u}$. Such a theory is clearly not the conventional gradient elasticity because of the presence of additional field variables. The question we pose is as follows: what are the possible physical scenarios represented by the new

Lagrangian? In view of the coupling of $\lambda$ and $\overline{\boldsymbol{\gamma}}$ with $\nabla \mathbf{C}$, it is evident that whenever strain gradients develop, an unknown scalar field and a vector field also emerge. One is reminded of the phenomenon of flexoelectricity where polarization and electric potential occur due to strain gradients. Therefore, we may associate $\lambda$ and $\overline{\boldsymbol{\gamma}}$ with the electric potential and polarization respectively. We later find that such a Lagrangian corresponds to isotropic flexoelectric materials only and the driving term for the 1-form field depends on terms which are isochoric. Although there exists materials, some polymers to wit, where deformation is approximately isochoric and for which the present theory may be effective, it still remains to investigate the local conformal invariance of the volumetric part which is out of scope of the present study. Presently, by contracting the quantity $\nabla \mathbf{C} - \boldsymbol{\gamma} \mathbf{C}$ in various ways, flexoelectricity for general anisotropic cases, piezoelectricity and electrostriction are modeled. We also look into magnetoelastic effects in static cases where we apply a variant of Hodge decomposition to the 1-form field $\boldsymbol{\kappa}$ as: $\boldsymbol{\kappa} = \nabla \times \boldsymbol{\tau} + \overline{\boldsymbol{\kappa}}$, $\boldsymbol{\tau}$ being a pseudo-vector field and $\overline{\boldsymbol{\kappa}}$ a vector field. Associating $\boldsymbol{\tau}$ and $\overline{\boldsymbol{\kappa}}$ with the magnetic potential and magnetization respectively, we extend our theory to model flexomagnetism, piezomagnetism and magnetostriction. In summary, starting with the conventional Lagrangian of elasticity theory and borrowing certain concepts from Weyl geometry, we are able to formulate a conformal gauge theory of solids by means of which a broad class of electromechanical and magnetomechanical phenomena may be rationally modelled. In order to briefly demonstrate what our proposal can do, we present an analytical solution considering piezoelectricity for isochoric deformation cases.

The rest of this paper is arranged as follows. A brief recap of material and spatial gauge theories of solids is provided in Section 2 which includes minimal replacement and minimal coupling constructs and an interpretation of the additional field variables. A gauge theory of solids with conformal symmetry is presented in Section 3, which contains discussions on conformal symmetries in classical elasticity theory, local conformal symmetries, the relation of the minimal replacement construct with Weyl geometry and minimal coupling. This section also provides a detailed derivation of the Euler-Lagrange equations from Hamilton's principle using the modified Lagrangian. Section 4 discusses the associations of the additional field variables with polarization and electric potential in the context of flexoelectricity. Flexoelectricity equations for non-centrosymmetric materials are discussed in Section 5. In Section 6, our model is extended to piezoelectricity. The electrostriction phenomenon is modeled with the conformal gauge theory in Section 7. Following this, we discuss a procedure to include magnetoelastic effects into our theory in Section 8. Equations for flexomagnetism are given in Section 9. Further on, we extend our theory to include piezomagnetism in Section 10. In Section 11, we derive the equations for magnetostriction. An illustrative example on piezoelectricity is presented in Section 12. Finally, we furnish a few concluding remarks in Section 13.

## 2 Material and spatial gauge theories of solids: a brief recap

For completeness, we provide a brief review of gauge theory of solids in line with Lagoudas and Edelen (1989) and Kadić and Edelen (1982). First, spatial and material symmetries in classical elasticity are discussed. The concept of local symmetries and minimal replacement are presented next, followed by a derivation of minimal coupling for the Lagrangian. Finally, a physical interpretation of additional field variables is given following Kadić and Edelen (1982).

### 2.1 Spatial and material symmetries in classical elasticity

The Lagrange density $L_0$ for a typical hyperelastic solid is given by:

$$L_0 = \frac{1}{2}\rho(\partial_t \chi)^T (\partial_t \chi) - U(\mathbf{C}) \tag{1}$$

Here $\chi$ is the deformation map, $\mathbf{C} = \mathbf{F}^T \mathbf{F}$ the right Cauchy-Green tensor with deformation gradient $\mathbf{F} = \frac{\partial \chi}{\partial \mathbf{X}}$. According to the principle of objectivity, $L_0$ is invariant under rigid body motions of the deformed configuration, i.e. under the transformations $`\chi = \mathbf{Q}\chi + \mathbf{b}$ and $`t = t$ with $\mathbf{Q} \in SO(3)$ and $\mathbf{b} \in T(3)$. Note that $\mathbf{Q}$ and $\mathbf{b}$ are constants. Similarly, the principle of frame indifference states that if the observer undergoes a rigid body motion, $L$ does not change. The corresponding transformation equations are: $`\mathbf{x} = \mathbf{Q}\mathbf{x} + \mathbf{b}$ and $`t = t$, where $\mathbf{x}$ is the spatial

coordinate. **Q** and **b** form the spatial symmetry group $G_s = SO(3) \rhd T(3)$ whose action is given by:

$$`\mathbf{\Psi} = \mathbf{A}\mathbf{\Psi} \tag{2}$$

where, $\mathbf{\Psi} = \begin{bmatrix} \chi^1 & \chi^2 & \chi^3 & 1 \end{bmatrix}^T$, $\mathbf{A} = \begin{bmatrix} \mathbf{Q} & \mathbf{b} \\ \mathbf{0} & 1 \end{bmatrix}$. The invariance of $L_0$ is stated as:

$$L_0\left(\partial_a `\mathbf{\Psi}\right) = L_0\left(\mathbf{A}\partial_a \mathbf{\Psi}\right) = L_0\left(\partial_a \mathbf{\Psi}\right) \tag{3}$$

Apart from the spatial symmetries, there also exists material symmetry in a solid body. Material symmetry deals with the invariance of $L_0$ with specific transformations of coordinates of the reference configuration. Lagoudas and Edelen (1989) prescribe these transformations as: $`X^A = R^A_B X^B + T^A$ and $`X^4 = X^4 + T^4$, where, $A, B \in \{1,2,3\}$, $\mathbf{R} \in SO(3)$ and $T \in T(1)$. The set of all these transformations forms a group given by: $G_m = \{SO(3) \rhd T(3)\} \times T(1)$, known as the material symmetry group. Finally, we have a total gauge group $G = G_m \times G_s$ under the action of which the Lagrange density ($L$) remains invariant.

### 2.2 Local symmetry and minimal replacement

Gauge theory investigates symmetries in the Lagrangian under the local action of a gauge group $G$, i.e. when parameters of $G$ depend on position and time. In the context of the spatial gauge group $G_s$, the local action is given by:

$$`\mathbf{\Psi}(X^a) = \mathbf{A}(X^a)\mathbf{\Psi}(X^a) \tag{4}$$

where $a \in \{1,2,3,4\}$. Here, time $t$ is incorporated as a fourth coordinate as: $X^4 = t$. As **A** is a function of space and time, we may write:

$$\partial_a `\mathbf{\Psi} = \mathbf{A}\partial_a \mathbf{\Psi} + (\partial_a \mathbf{A})\mathbf{\Psi} \tag{5}$$

Note that the derivatives of **A** have appeared on the right-hand side of equation (5). Substituting for $\partial_a \grave{\Psi}$ in the argument of $L_0$, we get:

$$L_0\left(\partial_a \grave{\Psi}\right) \neq L_0\left(\partial_a \Psi\right) \tag{6}$$

Thus $L_0$ is not invariant under local transformations. However, in gauge theory, one does seek invariance of $L_0$ under position and time dependent transformations. In order to find a solution to this problem, following Yang and Mills, Kadić and Edelen (1982) have modified the definition of the partial derivative operator $\partial_a$ to $\partial_a + \Gamma_a$, which is known as the gauge covariant derivative. The transformation of $\Gamma$ under the local action of $G$ is given by:

$$\grave{\Gamma}_a = \mathbf{A}\Gamma\mathbf{A}^{-1} - \left(\partial_a \mathbf{A}\right)\mathbf{A}^{-1} \tag{7}$$

The Lagrange density $L_0$ is constructed with the modified gradient $\partial_a \Psi + \Gamma_a \Psi$. Kadić and Edelen (1982) show that, under transformations of $\Psi$ and $\Gamma_a$ according to equations (4) and (7), $L_0$ remains invariant. More explicitly, writing the components of $\Gamma_a$, the deformation gradient is modified to:

$$B_a^i = \partial_a \chi^i + \phi_a^i + W_a^\alpha \gamma_{\alpha j}^i \chi^j \tag{8}$$

The modified Lagrange density may be written as:

$$L_0 = \frac{1}{2} \rho B_4^i B_4^i - U\left(\overline{\mathbf{C}}\right) \tag{9}$$

where, $\overline{\mathbf{C}} = \mathbf{B}^T \mathbf{B}$.

## 2.3 Minimal coupling

The Lagrange density ($L_0$) presented in equation (9) is algebraic in $\phi_a^i$ and $W_a^\alpha$. Therefore, unrealistic solutions arise when one uses this Lagrangian to obtain the Euler-Lagrange equations (Kadić and Edelen, 1982). In order to set this right, in line with Yang and Mills, Kadić and

Edelen (1982) add gauge invariant terms in the Lagrangian which are functions of the curvature associated with the connection 1-form $\Gamma_a$. The modified Lagrange density is written in the following form:

$$L = L_0 - s_1 L_\phi - s_2 L_w \tag{10}$$

where $L_\phi$ and $L_w$ are the additional gauge invariant terms, $s_1$ and $s_2$ material parameters and $L$ the minimally coupled Lagrange density. An explicit form of $L_\phi$ as prescribed by Kadić and Edelen (1982) is given below.

$$L_\phi = \frac{1}{2}\delta_{ij}\mathcal{D}^i_{ab}k^{am}k^{bn}\mathcal{D}^j_{mn} \tag{11}$$

Here, $k^{ab} = diag\left(-1,-1,-1,\frac{1}{y}\right)$ with $y$ being a material parameter. $\mathcal{D}^i_{ab}$ is written as:

$$\mathcal{D}^i_{ab} = \partial_a \phi^i_b - \partial_b \phi^i_a + \gamma^i_{\alpha j}\left(W^\alpha_a \phi^j_b - W^\alpha_b \phi^j_a + F^\alpha_{ab} x^j\right) \tag{12}$$

Similarly, $L_w$ is written as:

$$L_w = \frac{1}{2}C_{\alpha\beta}F^\alpha_{ab}g^{am}g^{bn}F^\beta_{mn} \tag{13}$$

$g^{\alpha\beta} = diag\left(-1,-1,-1,\frac{1}{\zeta}\right)$ and $\zeta$ is a material parameter. The expression for $F^\alpha_{ab}$ is (Kadić and Edelen 1982):

$$F^\alpha_{ab} = \partial_a W^\alpha_b - \partial_b W^\alpha_a + C^\alpha_{\beta\gamma}W^\beta_\alpha W^\gamma_b \tag{14}$$

## 2.4 Interpretation of compensating field variables

Initially Kadić and Edelen (1982) interpreted the compensating field variable $\phi^i_a$ and $W^\alpha_b$ arising from homogeneity-breaking of the action of groups $T(3)$ and $SO(3)$ as being associated with dislocations and disclinations respectively. However, at a later stage, exploiting spatial and material gauge symmetries of the Lagrangian, Lagoudas and Edelen (1989) derived spatial and material gauge connections, with which they constructed spatial and material gauge torsion and curvature. While spatial torsion and curvature are associated with microcracks and microrotations, material torsion and curvature are taken to represent dislocations and disclinations respectively.

## 3 Gauge theory of solids with conformal symmetry

We now consider a gauge theory of solids with conformal symmetry. First, a discussion on conformal symmetries in classical elasticity theory is presented. We show how the isochoric part of the Lagrange density loses its invariance under a local conformal transformation. Then, drawing an analogy with Weyl's geometry, we propose a minimal replacement construct of the covariant derivative to restore the invariance. Minimal coupling is presented next. Using Hamilton's principle, we derive the Euler-Lagrange equations. At this stage, we investigate the physical aspects that these equations could reveal. We observe that our equations are closely related to the governing equations of flexoelectricity, an electromechanical coupling phenomenon.

### 3.1 Conformal symmetries in classical elasticity theory

Distance between two infinitesimally close points in the reference configuration and their distance after deformation may be computed using the following expressions.

$$(dS)^2 = G_{\mu\nu} dX^\mu dX^\nu \tag{15}$$

$$(ds)^2 = g_{\mu\nu} dx^\mu dx^\nu = \left(\mathbf{F}^\mathbf{T} \mathbf{g} \mathbf{F}\right)_{\mu\nu} dX^\mu dX^\nu = C_{\mu\nu} dX^\mu dX^\nu \tag{16}$$

$dS$ and $ds$ are infinitesimal distances in the reference and current configurations respectively, $\mathbf{F}$ the deformation gradient, $\mathbf{G}$ and $\mathbf{g}$ metrics in the reference and current configurations respectively and $\mathbf{C}$ the right Cauchy-Green tensor ($\mathbf{g}$ pulled back to the reference configuration by the deformation). We may consider the geometry of the reference configuration as Euclidean;

in that case $\mathbf{G}$ is a Euclidean metric. Clearly, one may envision $\mathbf{C}$ as a metric by which infinitesimal distances in the current configuration can be computed using coordinates assigned to the reference configuration. Note that the geometry described by the metric $\mathbf{C}$ need not be Euclidean.

A conformal transformation of the metric $\mathbf{C}$ is defined by:

$$\grave{\mathbf{C}} = e^f \mathbf{C} \tag{17}$$

where $f$ is a real constant. This implies that $\mathbf{C}$ is scaled uniformly over the entire body. The dynamics of the body is governed by an action functional which involves a Lagrangian. Considering static deformation, the Lagrange density ($\mathcal{L}$) can be written as follows:

$$\mathcal{L} = -\Psi \tag{18}$$

where $\Psi$ denotes the strain energy density. Typically, $\Psi$ can be divided into the volumetric ($\Psi_{vol}$) and isochoric ($\Psi_{ic}$) parts as:

$$\Psi = \Psi_{vol} + \Psi_{ic} \tag{19}$$

Only $\Psi_{vol}$ is affected by the conformal transformation of $\mathbf{C}$ (equation 17), while $\Psi_{ic}$ remains unaffected. We choose $\Psi_{vol}$ and $\Psi_{ic}$ in the following way:

$$\Psi_{vol} = c_1 \left[ (\det \mathbf{C})^{\frac{1}{2}} - 1 \right]^2 \tag{20}$$

$$\Psi_{ic} = c_2 C_{\alpha\beta} G^{\alpha\mu} G^{\beta\nu} C_{\mu\nu} (\det \mathbf{C})^{-\frac{2}{3}} + c_3 \nabla_\alpha C_{\mu\nu} G^{\alpha\beta} G^{\mu\gamma} G^{\nu\kappa} \nabla_\beta C_{\gamma\kappa} (\det \mathbf{C})^{-\frac{2}{3}} \tag{21}$$

Here $\nabla$ denotes the covariant derivative. Note that we have included a gradient term (involving $\mathbf{C}$) in $\Psi_{ic}$. Under the conformal transformation of $\mathbf{C}$, one verifies that:

$$\grave{\Psi}_{vol} = c_1 \left[ e^{\frac{3f}{2}} (\det \mathbf{C})^{\frac{1}{2}} - 1 \right]^2 \neq \Psi_{vol} \tag{22}$$

$$\grave{\Psi}_{ic} = \Psi_{ic} \tag{23}$$

Therefore, conformal symmetry exists only in the isochoric part of the classical Lagrange density.

## 3.2 Local conformal transformation

If the scale factor is made to depend on position, i.e. $f = f(\mathbf{X})$, then under the transformation $`\mathbf{C} = e^{f(\mathbf{X})}\mathbf{C}$, $\Psi_{vol}$ and $\Psi_{ic}$ transform as:

$$`\Psi_{vol} \neq \Psi_{vol} \tag{24}$$

$$`\Psi_{ic} = c_2 C_{\alpha\beta} G^{\alpha\mu} G^{\beta\nu} C_{\mu\nu} (\det \mathbf{C})^{-\frac{2}{3}}$$
$$+ c_3 \nabla_\alpha \left(e^{f(\mathbf{X})} C_{\mu\nu}\right) G^{\alpha\beta} G^{\mu\gamma} G^{\nu\kappa} \nabla_\beta \left(e^{f(\mathbf{X})} C_{\gamma\kappa}\right) (\det \mathbf{C})^{-\frac{2}{3}} \neq \Psi_{ic} \tag{25}$$

In this case, the invariance of $\Psi_{ic}$ is also lost. This is because of the presence of $\nabla \mathbf{C}$ in $\Psi_{ic}$.

## 3.3 Relationship with Weyl geometry

We look for a way to preserve the invariance of $\Psi_{ic}$ under a local conformal transformation of $\mathbf{C}$. For this, we make use of certain concepts from Weyl geometry. Weyl tried to formulate a unified theory of gravitation and electricity (Weyl, 1918). In Einstein's theory of general relativity, space-time continuum is described by Riemann geometry. Here, the length of a vector remains unchanged under parallel transport. In Riemann geometry, the covariant derivative of the metric tensor is zero.

$$\nabla_\alpha g_{\mu\nu} = 0 \tag{26}$$

To incorporate the electrical field, Weyl assumed a new geometry of the space-time continuum, today known as Weyl geometry. He introduced the concept of a conformal transformation which involves a scaling of the metric as: $`\mathbf{g} = e^{f(\mathbf{X})}\mathbf{g}$. Weyl geometry is a generalization of Riemann geometry wherein the length of a vector does not remain unchanged upon parallel transport. Here an additional 1-form field **γ** is introduced with respect which we have a new metric compatibility condition:

$$\nabla_\alpha g_{\mu\nu} = \gamma_\alpha g_{\mu\nu} \tag{27}$$

This condition remains invariant under a group of transformations, known as Weyl's transformations.

$$`\mathbf{g} = e^f \mathbf{g} \tag{28}$$

$$`\mathbf{\gamma} = \mathbf{\gamma} + df \tag{29}$$

When $\mathbf{\gamma} = \mathbf{0}$, we recover the Riemann condition given by equation (26). We refer to the manifold as Weyl's integrable manifold when $\mathbf{\gamma} = d\lambda$, i.e. an exact 1-form. Relation between the metric tensor and connection may be obtained from equation (27) as (see Romero *et al*., 2012):

$$\tilde{\Gamma}^\alpha_{\mu\nu} = \Gamma^\alpha_{\mu\nu} - \frac{1}{2} g^{\alpha\beta} \left[ g_{\beta\mu}\gamma_\nu + g_{\beta\nu}\gamma_\mu - g_{\mu\nu}\gamma_\beta \right] \tag{30}$$

Here, $\Gamma$ and $\tilde{\Gamma}$ denote the Levi-Civita and Weyl connections respectively. Interpreting **C** as a metric, if one wishes to work with Riemann geometry in continuum mechanics, then $\nabla \mathbf{C}$ has to be zero, which is the Riemann condition. This can be achieved by incorporating a gradient term in the Lagrange density proportional to: $\nabla_\alpha C_{\mu\nu} G^{\alpha\beta} G^{\mu\gamma} G^{\nu\kappa} \nabla_\beta C_{\gamma\kappa}$. In the present work, we try to impose Weyl condition (equation 27) through the Lagrangian and see its consequences.

### 3.4 Minimal replacement

Following Weyl's condition, we modify the definition of the covariant derivative $\nabla$ to a gauge covariant derivative $\bar{\nabla}$ as:

$$\bar{\nabla} = \nabla - \mathbf{\gamma} \tag{31}$$

Now, using Weyl's transformations (equations 17 and 29), we may write the following identity.

$$\bar{\nabla}`\mathbf{C} = \nabla`\mathbf{C} - `\gamma`\mathbf{C} = \nabla\left(e^{f(\mathbf{X})}\mathbf{C}\right) - \left[\gamma + df\right]\left(e^{f(\mathbf{X})}\mathbf{C}\right)$$
$$= e^{f(\mathbf{X})}\left(\nabla\mathbf{C} - \gamma\mathbf{C}\right) = e^{f(\mathbf{X})}\bar{\nabla}\mathbf{C} \tag{32}$$

The attendant transformations of $\Psi_{vol}$ and $\Psi_{ic}$ reveal that:

$$`\Psi_{vol} \neq \Psi_{vol} \tag{33}$$

$$`\Psi_{ic} = \Psi_{ic} \tag{34}$$

Thus $\Psi_{ic}$ is invariant under the transformations given by equations (17) and (29), provided that the Lagrangian is constructed via the minimally replaced covariant derivative.

## 3.5 Minimal coupling

We now add gauge invariant terms based on curvature associated with the connection 1-forms to the Lagrangian. Decomposing $\gamma$ to its exact ($\gamma_e$) and anti-exact ($\bar{\gamma}$) parts, we may write:

$$\gamma = \gamma_e + \bar{\gamma} = d\lambda + \bar{\gamma} \tag{35}$$

Here, $\lambda$ is a 0-form field whose exterior derivative is the exact part of $\gamma$. One may construct a gauge invariant quantity from $\gamma$ in the following way.

$$d\gamma = d^2\lambda + d\bar{\gamma} = d\bar{\gamma} = \mathbf{Z} \tag{36}$$

Note that $\mathbf{Z}$ is a 2-form. Using $\mathbf{Z}$, we write the minimal coupling term in the Lagrange density as:

$$\Psi_{MC} = \frac{c_4}{4} Z_{ik} G^{ip} G^{kq} Z_{pq} \tag{37}$$

Using equations (20), (21) and (37), the minimally coupled $\Psi$ thus becomes:

$$\Psi = c_1\left[(\det \mathbf{C})^{\frac{1}{2}} - 1\right]^2 + c_2 C_{\alpha\beta} G^{\alpha\mu} G^{\beta\nu} C_{\mu\nu} (\det \mathbf{C})^{-\frac{2}{3}}$$

$$+ c_3 \left[\nabla_\alpha C_{\mu\nu} - (d\lambda + \overline{\gamma})_\alpha C_{\mu\nu}\right] G^{\alpha\beta} G^{\mu\gamma} G^{\nu\kappa} \left[\nabla_\beta C_{\gamma\kappa} - (d\lambda + \overline{\gamma})_\beta C_{\gamma\kappa}\right](\det \mathbf{C})^{-\frac{2}{3}} \quad (38)$$

$$+ \frac{c_4}{4} Z_{ik} G^{ip} G^{kq} Z_{pq}$$

In equation (38), the volumetric part of $\Psi$ changes under $\mathbf{C}' = e^f \mathbf{C}$ whereas the isochoric part remains invariant. One may as well try to formulate a gauge theory where $\Psi_{vol}$ will remain invariant along with $\Psi_{ic}$, but that is not our current concern. In order to use $\Psi$ in Hamilton's principle, we use gauge fixing where the gauge parameter $f$ is set to zero.

### 3.6 Hamilton's principle and Euler-Lagrange equations

Considering quasi-static deformation, Hamilton's principle may be stated as:

$$\int_{t_1}^{t_2} \left[\int_\Omega \delta \mathcal{L} dV + \int_\Omega f_\alpha \delta u_\alpha \, dV + \int_{\partial\Omega} \overline{t}_\alpha \delta u_\alpha \, dA\right] dt = 0, \qquad t_1 < t_2 \quad (39)$$

Note that, for simplicity in the exposition, we consider the geometry of the reference configuration to be Euclidean (i.e. the reference metric tensor is identity and the connection coefficients are zero) and write the Euler-Lagrange equations in Cartesian coordinates. The Lagrange density can be written as:

$$\mathcal{L} = -\Psi = -\Psi_{vol} - \Psi_{ic} - \Psi_{MC} \quad (40)$$

The volumetric and the isochoric parts of the stored energy density are given by:

$$\Psi_{vol} = c_1 \left[(\det \mathbf{C})^{\frac{1}{2}} - 1\right]^2 \quad (41)$$

$$\Psi_{ic} = c_2 C_{\mu\nu} C_{\mu\nu} (\det \mathbf{C})^{-\frac{2}{3}} + c_3 \left[\partial_\alpha C_{\mu\nu} - (\partial_\alpha \lambda + \overline{\gamma}_\alpha) C_{\mu\nu}\right]\left[\partial_\alpha C_{\mu\nu} - (\partial_\alpha \lambda + \overline{\gamma}_\alpha) C_{\mu\nu}\right](\det \mathbf{C})^{-\frac{2}{3}}$$

$$(42)$$

The minimal coupling term $\Psi_{MC}$ can be written as:

$$\Psi_{MC} = \frac{c_4}{4}\left(\partial_k \bar{\gamma}_i - \partial_i \bar{\gamma}_k\right)\left(\partial_k \bar{\gamma}_i - \partial_i \bar{\gamma}_k\right) \tag{43}$$

We now write the variation of $\Psi_{vol}$ as:

$$\delta\psi_{vol} = c_1 (\det \mathbf{C})^{\frac{1}{2}} \left[(\det \mathbf{C})^{\frac{1}{2}} - 1\right] C^{-1}_{\mu\nu} \delta C_{\mu\nu} \tag{44}$$

In the above equation, we have made use of the identity:

$$\delta(\det \mathbf{C}) = (\det \mathbf{C}) C^{-1}_{\mu\nu} \delta C_{\mu\nu} \tag{45}$$

Similarly, the variation of $\Psi_{ic}$ may be expressed as:

$$\delta\psi_{ic} = 2(\det \mathbf{C})^{-\frac{2}{3}} \left[ c_2 \left( C_{\mu\nu} - \frac{1}{3} C_{\alpha\beta} C_{\alpha\beta} C^{-1}_{\mu\nu} \right) \delta C_{\mu\nu} + c_3 \left(\partial_\alpha C_{\mu\nu} - \gamma_\alpha C_{\mu\nu}\right) \partial_\alpha \left(\delta C_{\mu\nu}\right) \right.$$
$$- c_3 \left(\partial_\alpha C_{\mu\nu} - \gamma_\alpha C_{\mu\nu}\right) \delta(\partial_\alpha \lambda + \bar{\gamma}_\alpha) C_{\mu\nu} - c_3 \left(\partial_\alpha C_{\mu\nu} - \gamma_\alpha C_{\mu\nu}\right) \gamma_\alpha \delta C_{\mu\nu} \tag{46}$$
$$\left. - \frac{1}{3}\left(\partial_\alpha C_{\beta\gamma} - \gamma_\alpha C_{\beta\gamma}\right)\left(\partial_\alpha C_{\beta\gamma} - \gamma_\alpha C_{\beta\gamma}\right) C^{-1}_{\mu\nu} \delta C_{\mu\nu} \right]$$

We may also write the variation of $\Psi_{MC}$ in the following manner.

$$\delta\Psi_{MC} = c_4 \left(\partial_\beta \bar{\gamma}_\alpha - \partial_\alpha \bar{\gamma}_\beta\right) \partial_\beta \left(\delta \bar{\gamma}_\alpha\right) \tag{47}$$

Substituting equations (44), (46) and (47) in Hamilton's principle and applying the divergence theorem, we obtain:

$$\int_{t_1}^{t_2} \int_\Omega \left[ S_{\mu\nu} \delta(C_{\mu\nu}/2) + P\delta\lambda + Q_\alpha \delta\bar{\gamma}_\alpha \right] dV dt$$
$$+ \int_{t_1}^{t_2} \int_{\partial\Omega} 2(\det \mathbf{C})^{-\frac{2}{3}} \left[ c_3 \left(\partial_\alpha C_{\mu\nu} - \gamma_\alpha C_{\mu\nu}\right) n_\alpha \delta C_{\mu\nu} - c_3 \left(\partial_\alpha C_{\mu\nu} - \gamma_\alpha C_{\mu\nu}\right) C_{\mu\nu} n_\alpha \delta\lambda \right] dA dt$$
$$+ \int_{t_1}^{t_2} \int_{\partial\Omega} c_4 \left(\partial_\beta \bar{\gamma}_\alpha - \partial_\alpha \bar{\gamma}_\beta\right) n_\beta \left(\delta\bar{\gamma}_\alpha\right) dA dt - \int_{t_1}^{t_2} \left[ \int_\Omega f_\alpha \delta u_\alpha dV + \int_{\partial\Omega} \bar{t}_\alpha \delta u_\alpha dA \right] dt = 0$$

$$\tag{48}$$

where,

$$S_{\mu\nu} = 2c_1 (\det \mathbf{C})^{\frac{1}{2}} \left[ (\det \mathbf{C})^{\frac{1}{2}} - 1 \right] C_{\mu\nu}^{-1} + 4c_2 (\det \mathbf{C})^{-\frac{2}{3}} \left( C_{\mu\nu} - \frac{1}{3} C_{\alpha\beta} C_{\alpha\beta} C_{\mu\nu}^{-1} \right)$$
$$- 4c_3 \partial_\alpha \left[ (\det \mathbf{C})^{-\frac{2}{3}} \left( \partial_\alpha C_{\mu\nu} - \gamma_\alpha C_{\mu\nu} \right) \right] \quad (49)$$
$$- 4c_3 (\det \mathbf{C})^{-\frac{2}{3}} \left[ \gamma_\alpha \left( \partial_\alpha C_{\mu\nu} - \gamma_\alpha C_{\mu\nu} \right) + \frac{1}{3} C_{\mu\nu}^{-1} \left( \partial_\alpha C_{\beta\gamma} - \gamma_\alpha C_{\beta\gamma} \right) \left( \partial_\alpha C_{\beta\gamma} - \gamma_\alpha C_{\beta\gamma} \right) \right]$$

$$P = c_3 \partial_\alpha \left[ (\det \mathbf{C})^{-\frac{2}{3}} \left( \partial_\alpha \left( C_{\mu\nu} C_{\mu\nu} \right) - 2\gamma_\alpha C_{\mu\nu} C_{\mu\nu} \right) \right] \quad (50)$$

and

$$Q_\alpha = -c_3 (\det \mathbf{C})^{-\frac{2}{3}} \left[ \partial_\alpha \left( C_{\mu\nu} C_{\mu\nu} \right) - 2\gamma_\alpha C_{\mu\nu} C_{\mu\nu} \right] - c_4 \partial_\beta \left( \partial_\beta \bar{\gamma}_\alpha - \partial_\alpha \bar{\gamma}_\beta \right) \quad (51)$$

The first term of the integrand in equation (48) may be simplified to:

$$\frac{1}{2} S_{\mu\nu} \delta C_{\mu\nu} = \frac{1}{2} S_{\mu\nu} \delta \left( F_{\alpha\mu} F_{\alpha\nu} \right) = \frac{1}{2} \left( S_{\mu\nu} F_{\alpha\nu} \delta F_{\alpha\mu} + S_{\mu\nu} F_{\alpha\mu} \delta F_{\alpha\nu} \right) = S_{\mu\nu} F_{\alpha\nu} \delta F_{\alpha\mu} \quad (52)$$

Using equation (52), we write the first term of equation (48) as:

$$\int_{t_1}^{t_2} \int_\Omega S_{\mu\nu} F_{\alpha\nu} \delta F_{\alpha\mu} dV dt = \int_{t_1}^{t_2} \int_\Omega S_{\mu\nu} F_{\alpha\nu} \partial_\mu \left( \delta u_\alpha \right) dV dt$$
$$= -\int_{t_1}^{t_2} \int_\Omega \partial_\mu \left( F_{\alpha\nu} S_{\nu\mu} \right) \delta u_\alpha dV dt + \int_{t_1}^{t_2} \int_{\partial\Omega} F_{\alpha\nu} S_{\nu\mu} n_\mu \delta u_\alpha dA dt \quad (53)$$

Substituting equation (53) in equation (48), we get:

$$-\int_{t_1}^{t_2} \int_\Omega \partial_\mu \left( F_{\alpha\nu} S_{\nu\mu} \right) \delta u_\alpha dV dt + \int_{t_1}^{t_2} \int_{\partial\Omega} F_{\alpha\nu} S_{\nu\mu} n_\mu \delta u_\alpha dA dt + \int_{t_1}^{t_2} \int_\Omega \left( P \delta\lambda + Q_\alpha \delta\bar{\gamma}_\alpha \right) dV dt$$
$$+ \int_{t_1}^{t_2} \int_{\partial\Omega} 2 (\det \mathbf{C})^{-\frac{2}{3}} \left[ c_3 \left( \partial_\alpha C_{\mu\nu} - \gamma_\alpha C_{\mu\nu} \right) n_\alpha \delta C_{\mu\nu} - c_3 \left( \partial_\alpha C_{\mu\nu} - \gamma_\alpha C_{\mu\nu} \right) C_{\mu\nu} n_\alpha \delta\lambda \right] dA dt \quad (54)$$
$$+ \int_{t_1}^{t_2} \int_{\partial\Omega} c_4 \left( \partial_\beta \bar{\gamma}_\alpha - \partial_\alpha \bar{\gamma}_\beta \right) n_\beta \delta\bar{\gamma}_\alpha dA dt - \int_{t_1}^{t_2} \left[ \int_\Omega f_\alpha \delta u_\alpha dV + \int_{\partial\Omega} \bar{t}_\alpha \delta u_\alpha dA \right] dt = 0$$

As $\delta \mathbf{u}$, $\delta\lambda$ and $\delta\bar{\gamma}$ are independent and arbitrary, we may write the Euler-Lagrange equations (on $\Omega$) as:

$$\partial_\mu (\mathbf{FS})_{\alpha\mu} + f_\alpha = 0 \tag{55}$$

$$Q_\alpha = 0 \tag{56}$$

$$P = 0 \tag{57}$$

The natural boundary conditions (on $\partial\Omega$) may be written as:

$$(\mathbf{FS})_{\alpha\mu} n_\mu = \bar{t}_\alpha \tag{58}$$

$$c_3 \left( \partial_\alpha C_{\mu\nu} - \gamma_\alpha C_{\mu\nu} \right) n_\alpha = 0 \tag{59}$$

$$c_4 \left( \partial_\beta \bar{\gamma}_\alpha - \partial_\alpha \bar{\gamma}_\beta \right) n_\beta = 0 \tag{60}$$

Either $\mathbf{C}$ is specified or $c_3 \left( \partial_\alpha C_{\mu\nu} - \gamma_\alpha C_{\mu\nu} \right) C_{\mu\nu} n_\alpha = 0$ on $\partial\Omega$ \tag{61}

## 4 Interpretation of additional field variables: emergence of flexoelectricity

In the Euler-Lagrange equations we just derived, there are three independent field variables describing the model kinematics; these are the displacement $\mathbf{u}$, the exact ($\lambda$) and the anti-exact ($\bar{\gamma}$) parts of the 1-form field $\gamma$. Clearly, the state of the body described by equations (55) through (57) departs from classical elasticity theory. Therefore, the question that naturally arises concerns the nature of physical phenomena which may be described by the new Lagrange density. In order to find an answer to this question, we notice that, $\lambda$ and $\bar{\gamma}$ are coupled with $\nabla \mathbf{C}$ in the new Lagrangian. Thus, owing to strain gradients, a scalar field $\lambda$ and a vector field $\bar{\gamma}$ are also developed. This brings to focus the phenomenon of flexoelectricity which is a coupling between strain gradient and polarization (Maranganti *et al.*, 2006, Majdoub *et al.*, 2008, Roy and Roy, 2018). In a flexoelectric material, high strain gradients cause the centers of positive and negative charges to displace from each other in a unit cell, thereby creating a dipole moment whose macroscopic density is known as polarization (Maranganti *et al.*, 2006). This electromechanical phenomenon is present in all dielectric materials and its effect is pronounced in the nanoscale where high strain gradients are common.

If an electric field vector ($\mathbf{E}$) induces polarization ($\mathbf{P}$) in the body, one may define an electric displacement vector ($\mathbf{D}$) as:

$$\mathbf{D} = \varepsilon_0 \mathbf{E} + \mathbf{P} \tag{62}$$

where $\varepsilon_0$ is the permittivity of free space. $\mathbf{E}$ may be obtained as the gradient of a scalar field known as the electric potential ($\varphi$): $E = -\nabla \varphi$. However, for many electromechanical phenomena, $\mathbf{D}$ depends on strain and its gradient in the presence of mechanical loading and equation (62) is modified with corresponding terms. For such a modified definition of $\mathbf{D}$ in the case of piezoelectricity, refer to Tiersten, 2013. In our case, the coupling amongst $\lambda$, $\bar{\gamma}$ and $\nabla \mathbf{C}$ arises due to the imposition of Weyl's condition through the Lagrangian. Keeping this in view, we interpret $\lambda = \varepsilon_0 \mu \varphi$ and $\bar{\gamma} = -\mu \mathbf{P}$, where $\mu$ is a proportionality constant used to make the units consistent. One may also envision $\gamma = -\mu \mathbf{D}$, where $\mathbf{D}$ is the electric displacement in the absence of mechanical loading. Solving the coupled system of equations (55) through (57), we can obtain $\mathbf{u}$, $\lambda$ and $\bar{\gamma}$. Electric potential and polarization fields may then be calculated using the relations: $\lambda = \varepsilon_0 \mu \varphi$ and $\bar{\gamma} = -\mu \mathbf{P}$. A close observation of our equation reveals that the 1-form field $\gamma$ is directly coupled with terms which are isochoric and not volumetric. This is a consequence of considering only local conformal symmetry of the isochoric part of the Lagrangian. Although it will be an interesting study to consider the local conformal invariance of the volumetric part also, presently we keep it as a future study and concentrate on only those materials (e.g. some polymers) for which isochoric parts plays a major role in the generation of electric field and polarization. The rest of the paper follows this viewpoint.

We now contrast our equilibrium equations with the presently known equations of flexoelectricity for centrosymmetric materials (Table 1). Note that, unlike our finite-deformation setting, the currently available flexoelectricity equations are written in the small deformation setup. For details of the constitutive equations for flexoelectricity, we refer to Maranganti *et al*. (2006).

Table 1: Available equations for flexoelectricity in isotropic materials: (a), (b) and (c); equations from our conformal gauge theory: (d), (e) and (f)

| | |
|---|---|
| $\nabla \cdot \hat{\boldsymbol{\sigma}} + \mathbf{f} = 0$ with $\hat{\boldsymbol{\sigma}} = \boldsymbol{\sigma} - \nabla \cdot \tilde{\boldsymbol{\sigma}}$ (a) | $\nabla \cdot (\mathbf{FS}) + \mathbf{f} = \mathbf{0}$ (d) |
| $-\varepsilon_0 \nabla^2 \varphi + \nabla \cdot \mathbf{P} = 0$ (b) | $\nabla \cdot \left[ (\det \mathbf{C})^{-\frac{2}{3}} \left( \nabla (\mathbf{C}:\mathbf{C}) - 2\gamma (\mathbf{C}:\mathbf{C}) \right) \right] = 0$ (e) |
| $(d_{44} - f_{12}) \nabla^2 \mathbf{u} + (d_{12} + d_{44} - 2f_{44}) \nabla \nabla \cdot \mathbf{u}$ $+ (b_{44} + b_{77}) \nabla^2 \mathbf{P} + (b_{12} + b_{44} - b_{77}) \nabla \nabla \cdot \mathbf{P}$ (c) $-a\mathbf{P} - \nabla \varphi + \mathbf{E}^0 = \mathbf{0}$ | $c_3 (\det \mathbf{C})^{-\frac{2}{3}} \left[ \nabla (\mathbf{C}:\mathbf{C}) - 2\gamma (\mathbf{C}:\mathbf{C}) \right]$ $+ c_4 \nabla \cdot \left[ \nabla \gamma - (\nabla \gamma)^T \right] = \mathbf{0}$ (f) |

In Table 1, $d_{12}$, $d_{44}$, $f_{12}$, $f_{44}$, $b_{12}$, $b_{44}$, $b_{77}$ and $a$ are material parameters and $\tilde{\boldsymbol{\sigma}}$ the higher order stress tensor (Maranganti *et al*., 2006).

In the classical formulation of flexoelectricity (Majdoub *et al*., 2008), the electric enthalpy density defined by Toupin is used. It is composed of two parts: energy density of deformation and polarization $W^L$ and another part corresponding to electric field and polarization.

$$H = W^L (\boldsymbol{\varepsilon}, \mathbf{P}, \nabla \nabla \mathbf{u}, \nabla \mathbf{P}) - \frac{1}{2} \varepsilon_0 \partial_i \varphi \partial_i \varphi + \partial_i \varphi P_i \tag{63}$$

One observes that the electric potential $\varphi$ has no direct coupling with strain $\boldsymbol{\varepsilon}$ in equation (63). Last two terms in equation (63) are written such that it leads to Gauss law for dielectrics.

$$\nabla \cdot \mathbf{D} = 0 \tag{64}$$

with $\mathbf{D}$ given by equation (62). However, in our formulation, no such assumption is made and therefore, we have a direct coupling between $\varphi$ and $\mathbf{C}$.

## 5. Flexoelectricity equations for non-centrosymmetric solids

We have observed that the new Lagrange density in Section 4 could be potentially useful as a model for flexoelectricity in centrosymmetric materials. However, for non-centrosymmetric materials, we need to generalize such a model. This may be achieved through a contraction of the quadratic of the difference $\nabla \mathbf{C} - \gamma \mathbf{C}$ using a sixth order constitutive tensor. We thus construct a gauge invariant term in the following way:

$$\Psi_{fle} = h_{\alpha\mu\nu pqr}\left(\partial_\alpha C_{\mu\nu} - \gamma_\alpha C_{\mu\nu}\right)\left(\partial_p C_{qr} - \gamma_p C_{qr}\right)\left(\det \mathbf{C}\right)^{-\frac{2}{3}} \tag{65}$$

Here $\mathbf{h}$ is a sixth order constitutive tensor. One may write the variation of $\Psi_{fle}$ as:

$$\begin{aligned}
\delta\Psi_{fle} &= h_{\alpha\mu\nu pqr}\left(\partial_\alpha \delta C_{\mu\nu} - \delta\gamma_\alpha C_{\mu\nu} - \gamma_\alpha \delta C_{\mu\nu}\right)\left(\partial_p C_{qr} - \gamma_p C_{qr}\right)\left(\det \mathbf{C}\right)^{-\frac{2}{3}} \\
&+ h_{pqr\alpha\mu\nu}\left(\partial_p C_{qr} - \gamma_p C_{qr}\right)\left(\partial_\alpha \delta C_{\mu\nu} - \delta\gamma_\alpha C_{\mu\nu} - \gamma_\alpha \delta C_{\mu\nu}\right)\left(\det \mathbf{C}\right)^{-\frac{2}{3}} \\
&- \frac{2}{3} h_{\alpha mn pqr} C^{-1}_{\mu\nu} \delta C_{\mu\nu}\left(\partial_\alpha C_{mn} - \gamma_\alpha C_{mn}\right)\left(\partial_p C_{qr} - \gamma_p C_{qr}\right)\left(\det \mathbf{C}\right)^{-\frac{2}{3}}
\end{aligned} \tag{66}$$

Using equation (66), we may now identify the additional terms that need to be added, i.e. $\mathbf{S}_{add,fle}$, $P_{add,fle}$ and $\mathbf{Q}_{add,fle}$ with $\mathbf{S}$, $P$ and $\mathbf{Q}$ respectively given by equations (49) through (51). The expression for $\mathbf{S}_{add,fle}$ is:

$$\begin{aligned}
\left(S_{add,fle}\right)_{\mu\nu} &= -2\partial_\alpha\left[\left(h_{\alpha\mu\nu pqr} + h_{pqr\alpha\mu\nu}\right)\left(\partial_p C_{qr} - \gamma_p C_{qr}\right)\left(\det \mathbf{C}\right)^{-\frac{2}{3}}\right] \\
&- 2\left(h_{\alpha\mu\nu pqr} + h_{pqr\alpha\mu\nu}\right)\gamma_\alpha\left(\partial_p C_{qr} - \gamma_p C_{qr}\right)\left(\det \mathbf{C}\right)^{-\frac{2}{3}} \\
&- \frac{4}{3} h_{\alpha mn pqr} C^{-1}_{\mu\nu}\left(\partial_\alpha C_{mn} - \gamma_\alpha C_{mn}\right)\left(\partial_p C_{qr} - \gamma_p C_{qr}\right)\left(\det \mathbf{C}\right)^{-\frac{2}{3}}
\end{aligned} \tag{67}$$

$P_{add,fle}$ may be written as:

$$P_{add,fle} = \partial_\alpha\left[\left(h_{\alpha\mu\nu pqr} + h_{pqr\alpha\mu\nu}\right) C_{\mu\nu}\left(\partial_p C_{qr} - \gamma_p C_{qr}\right)\left(\det \mathbf{C}\right)^{-\frac{2}{3}}\right] \tag{68}$$

Similarly, the expression for $\mathbf{Q}_{add,fle}$ may be written as follows:

$$\left(Q_{add,fle}\right)_\alpha = -\left(h_{\alpha\mu\nu pqr} + h_{pqr\alpha\mu\nu}\right)C_{\mu\nu}\left(\partial_p C_{qr} - \gamma_p C_{qr}\right)\left(\det \mathbf{C}\right)^{-\frac{2}{3}} \tag{69}$$

## 6 Extension to piezoelectricity

We now discuss a possible model for finite-deformation piezoelectricity on the lines of the new Lagrangian density. Piezoelectricity is a linear electromechanical coupling between strain and polarization and occurs in non-centrosymmetric materials only. Many naturally occurring materials as well as biological substances show piezoelectric effect; for example, quartz, Rochelle salt, silk, DNA, viral proteins, bones, enamel etc. These apart, certain ceramics and polymers also show this effect, e.g. lead zirconate titanate (PZT), Barium titanate ($BaTiO_3$), Polyvinylidene fluoride etc. Piezoelectricity is widely exploited in industrial applications, e.g. sensors and actuators, energy harvesting etc., as this effect is prominent in the macroscale.

In order to model piezoelectricity through the conformal gauge theory, we need to contract the gauge invariant quantity $\left(\nabla\mathbf{C} - \boldsymbol{\gamma}\mathbf{C}\right)\left(\det\mathbf{C}\right)^{-\frac{1}{3}}$ by a third order constitutive tensor. The additional term in the Lagrange density may be written as follows:

$$\Psi_{pe} = e_{\alpha\mu\nu}\left(\partial_\alpha C_{\mu\nu} - \gamma_\alpha C_{\mu\nu}\right)\left(\det\mathbf{C}\right)^{-\frac{1}{3}} \tag{70}$$

Here $\mathbf{e}$ is a third order constitutive tensor. In the case of small strain gradients, one may notice that $\Psi_{pe}$ can be approximated as: $\Psi_{pe} = -e_{\alpha\mu\nu}\gamma_\alpha C_{\mu\nu}\left(\det\mathbf{C}\right)^{-\frac{1}{3}}$. This term shows a close resemblance with the energy density in the standard piezoelectricity theory, $\Psi_{classical,pe}$ (applicable in the small-deformation setting). It is typically written as: $\Psi_{classical,pe} = \tilde{e}_{\alpha\mu\nu}P_\alpha\varepsilon_{\mu\nu}$, where $\tilde{\mathbf{e}}$ is a constant third order tensor and $\boldsymbol{\varepsilon}$ the small strain matrix. Therefore, we may expect to model finite deformation piezoelectricity with the proposed theory.

The variation of $\Psi_{pe}$ is given by:

$$\delta\Psi_{pe} = e_{\alpha\mu\nu}\left(\partial_\alpha \delta C_{\mu\nu} - \delta\gamma_\alpha C_{\mu\nu} - \gamma_\alpha \delta C_{\mu\nu}\right)(\det \mathbf{C})^{-\frac{1}{3}}$$
$$-\frac{1}{3}(\det \mathbf{C})^{-\frac{1}{3}} C_{\mu\nu}^{-1} \delta C_{\mu\nu} e_{pqr}\left(\partial_p C_{qr} - \gamma_p C_{qr}\right) \tag{71}$$

Piezoelectric equations of equilibrium may be obtained via addition of terms $\mathbf{S}_{add,pe}$, $P_{add,pe}$ and $\mathbf{Q}_{add,pe}$ with $\mathbf{S}$, $P$ and $\mathbf{Q}$ given respectively in equations (49) through (51). We may write $\mathbf{S}_{add,pe}$ as follows:

$$\left(S_{add,pe}\right)_{\mu\nu} = -2\partial_\alpha\left[e_{\alpha\mu\nu}(\det \mathbf{C})^{-\frac{1}{3}}\right] - 2(\det \mathbf{C})^{-\frac{1}{3}}\left[e_{\alpha\mu\nu}\gamma_\alpha + \frac{1}{3}C_{\mu\nu}^{-1} e_{pqr}\left(\partial_p C_{qr} - \gamma_p C_{qr}\right)\right] \tag{72}$$

$P_{add,pe}$ may be expressed as:

$$P_{add,pe} = \partial_\alpha\left[e_{\alpha\mu\nu} C_{\mu\nu}(\det \mathbf{C})^{-\frac{1}{3}}\right] \tag{73}$$

Finally, the expression for $\mathbf{Q}_{add,pe}$ is:

$$\left(Q_{add,pe}\right)_\alpha = -e_{\alpha\mu\nu} C_{\mu\nu}(\det \mathbf{C})^{-\frac{1}{3}} \tag{74}$$

Note that the strain gradient terms, which appear naturally in our formulation, have typically not been used with standard piezoelectricity models.

In Section 12, an assessment of the proposed governing equations and its comparison with the classical linear theory of piezoelectricity is given.

## 7 Modelling electrostriction

Electrostriction is an electromechanical coupling phenomenon encountered in all dielectric materials, wherein strain is coupled with the quadratic of polarization. The electrostrictive strain is given by $\varepsilon_{ij} = q_{ijkl} P_k P_l$ (Newnham *et al.* 1997), where $\mathbf{q}$ is a fourth order constitutive tensor and $\mathbf{P}$ the polarization vector. Although electrostriction exists in all dielectrics to an extent, its effect in most materials is barely perceptible, and therefore of no great practical use. However some materials show high electrostrictive coefficients, for example, relaxor ferroelectrics such as Lead Magnesium Niobate (Pirc *et al.*, 2004), Lead Lanthanum Zirconate Titanate etc., ceramics, Barium Titanate (Mason, 1948) and polymers: electrostrictive polyurethane (Ask *et al.*, 2012). Yang and Suo (1994) have theoretically studied fracture in actuators under electrical loading considering the electrostriction effect. Richards and Odegard (2010) have developed a thermodynamic framework for electrostrictive polymers. A review on energy harvesting from electrostrictive polymers is available in Lallart *et al.* (2012). Guo and Guo (2003) have theoretically predicted the possibility of giant electrostrictive deformation in carbon nanotubes.

In order to model electrostriction in the conformal gauge theoretic setup, we need to add an additional gauge invariant term in the Lagrange density. First note that, under the transformation $\grave{\mathbf{C}} = e^{f(\mathbf{X})}\mathbf{C}$, the terms $C_{pq}^{-1}\left(\partial_k C_{pq} - \gamma_k C_{pq}\right)$ and $\left(\partial_l C_{ij} - \gamma_l C_{ij}\right)(\det \mathbf{C})^{-\frac{1}{3}}$ remain invariant. Using these quantities, we may construct a gauge invariant term in the following way.

$$\Psi_{es} = q_{ijkl}\left(C_{pq}^{-1}\partial_k C_{pq} - 3\gamma_k\right)\left(\partial_l C_{ij} - \gamma_l C_{ij}\right)(\det \mathbf{C})^{-\frac{1}{3}} \tag{75}$$

where, **q** is a fourth order constitutive tensor. In cases where the strain gradient effect is negligible, $\Psi_{es}$ may be approximately written as: $\Psi_{es} \simeq 3q_{ijkl}\gamma_k\gamma_l C_{ij}(\det \mathbf{C})^{-\frac{1}{3}}$. This shows a close correspondence with the conventional electroctritive energy density (for small deformation), which is written as: $\Psi_{classical,es} = \tilde{q}_{ijkl} P_k P_l \varepsilon_{ij}$, where $\tilde{\mathbf{q}}$ is a constant fourth order tensor. We also note that, in our finite deformation electrostriction, the gradient of strain continues to play a significant role. Variation of $\Psi_{es}$ can be written as:

$$\begin{aligned}\delta\Psi_{es} = & \ q_{ijcd}\left(-C_{p\mu}^{-1}C_{qv}^{-1}\partial_\alpha C_{pq}\delta C_{\mu v} + C_{\mu v}^{-1}\partial_\alpha \delta C_{\mu v} - 3\delta\gamma_\alpha\right)\left(\partial_l C_{ij} - \gamma_l C_{ij}\right)(\det \mathbf{C})^{-\frac{1}{3}} \\ & + q_{\mu v l \alpha}\left(C_{pq}^{-1}\partial_l C_{pq} - 3\gamma_l\right)\left(\partial_\alpha \delta C_{\mu v} - \delta\gamma_\alpha C_{\mu v} - \gamma_\alpha \delta C_{\mu v}\right)(\det \mathbf{C})^{-\frac{1}{3}} \\ & - \frac{1}{3}q_{ijkl}C_{\mu v}^{-1}\delta C_{\mu v}\left(C_{pq}^{-1}\partial_k C_{pq} - 3\gamma_k\right)\left(\partial_l C_{ij} - \gamma_l C_{ij}\right)(\det \mathbf{C})^{-\frac{1}{3}}\end{aligned} \tag{76}$$

Using this, we may now write the extra terms $\mathbf{S}_{add,es}$, $P_{add,es}$ and $\mathbf{Q}_{add,es}$ that need to be added with $\mathbf{S}$, $P$ and $\mathbf{Q}$ respectively (see equations 49 through 51) in the equilibrium equations to account for electrostrictive effects. $\mathbf{S}_{add,es}$ is given as:

$$\begin{aligned}\left(S_{add,es}\right)_{\mu\nu} &= -2q_{ijal}C^{-1}_{p\mu}C^{-1}_{qv}\partial_{\alpha}C_{pq}\left(\partial_{l}C_{ij}-\gamma_{l}C_{ij}\right)\left(\det \mathbf{C}\right)^{-\frac{1}{3}} \\
&\quad -2q_{\mu\nu l\alpha}\left(C^{-1}_{pq}\partial_{l}C_{pq}-3\gamma_{l}\right)\gamma_{\alpha}\left(\det \mathbf{C}\right)^{-\frac{1}{3}} \\
&\quad -\frac{2}{3}q_{ijkl}C^{-1}_{\mu\nu}\left(C^{-1}_{pq}\partial_{k}C_{pq}-3\gamma_{k}\right)\left(\partial_{l}C_{ij}-\gamma_{l}C_{ij}\right)\left(\det \mathbf{C}\right)^{-\frac{1}{3}} \\
&\quad -2\partial_{\alpha}\left[q_{ijal}C^{-1}_{\mu\nu}\left(\partial_{l}C_{ij}-\gamma_{l}C_{ij}\right)\left(\det \mathbf{C}\right)^{-\frac{1}{3}}+q_{\mu\nu l\alpha}\left(C^{-1}_{pq}\partial_{l}C_{pq}-3\gamma_{l}\right)\left(\det \mathbf{C}\right)^{-\frac{1}{3}}\right]\end{aligned} \quad (77)$$

The expression for $P_{add,es}$ is as follows:

$$P_{add,es} = \partial_{\alpha}\left[3q_{ijal}\left(\partial_{l}C_{ij}-\gamma_{l}C_{ij}\right)\left(\det \mathbf{C}\right)^{-\frac{1}{3}}+q_{\mu\nu l\alpha}\left(C^{-1}_{pq}\partial_{l}C_{pq}-3\gamma_{l}\right)C_{\mu\nu}\left(\det \mathbf{C}\right)^{-\frac{1}{3}}\right] \quad (78)$$

Finally, we may write $\left(Q_{add,es}\right)_{\alpha}$ as:

$$\left(Q_{add,es}\right)_{\alpha} = -\left[3q_{ijal}\left(\partial_{l}C_{ij}-\gamma_{l}C_{ij}\right)+q_{\mu\nu l\alpha}\left(C^{-1}_{pq}\partial_{l}C_{pq}-3\gamma_{l}\right)C_{\mu\nu}\right]\left(\det \mathbf{C}\right)^{-\frac{1}{3}} \quad (79)$$

## 8 Modelling magnetoelasticity

We have discussed how various electromechanical coupling phenomena, e.g. flexoelectricity, piezoelectricity and electrostriction are derivable using the conformal gauge theory of solids. In all these cases, we consider only quasi-static conditions, i.e. where time derivatives are not taken into account. In the presence of time dependent motion of the body, the bound charges move and a magnetic field is generated along with the electric field. Apart from dynamic conditions, magnetism may be present even in static cases and many magnetomechanical coupling phenomena may also exist, viz. flexomagnetism, piezomagnetism, magnetostriction etc. In classical electromagnetism, electricity and magnetism are represented through components of an electromagnetic tensor in terms of which Maxwell's equations are written. We also note that

Maxwell's equations for electricity and magnetism are uncoupled in the absence of time dependent effects. In static cases, the electric field is represented as the gradient of a scalar potential and the magnetic field through the curl of a pseudo-vector potential.

In the following sections, we deal only with magnetostatic cases. However, in order to obtain a suitable decomposition of the additional 1-form field, we briefly touch upon the 4-dimensional scenario that involves time as the fourth coordinate. In the 4-dimensional setup, one may write Weyl's condition as: $\nabla_\alpha g_{\mu\nu} = \omega_\alpha g_{\mu\nu}$, where, $\alpha, \mu, \nu \in \{1,2,3,4\}$. Using the Hodge decomposition theorem, any $p$-form $\omega \in \Lambda^p(\mathcal{M})$ on a compact orientable smooth manifold $\mathcal{M}$ of dimension $n$, may be written as (Ivancevic and Ivancevic, 2008):

$$\kappa = d\alpha + \bar{\delta}\beta + \bar{\kappa} \tag{80}$$

Here, $\alpha \in \Lambda^{p-1}(\mathcal{M})$, $\beta \in \Lambda^{p+1}(\mathcal{M})$ and $\bar{\kappa} \in \Lambda^p(\mathcal{M})$. $d\alpha$, $\bar{\delta}\beta$ and $\bar{\kappa}$ respectively are the exact, coexact and harmonic parts of $\kappa$. Recall that $\Lambda^p$ is the space of differential forms of order $p$. The operators $\bar{\delta}: \Lambda^p(\mathcal{M}) \to \Lambda^{p-1}(\mathcal{M})$ are defined as follows:

$$\bar{\delta} = (-1)^{n(p+1)+1} * d * \tag{81}$$

where, $*$ is the Hodge star operator. In order to bring in the magnetic effect, we need to decompose the 1-form field in a similar manner. Writing equation (80) in component form, we have:

$$\begin{pmatrix} \kappa_1 \\ \kappa_2 \\ \kappa_3 \\ \kappa_4 \end{pmatrix} = \begin{pmatrix} \partial_1\alpha \\ \partial_2\alpha \\ \partial_3\alpha \\ \partial_4\alpha \end{pmatrix} - \begin{pmatrix} -(\partial_2\bar{\beta}_{12} + \partial_3\bar{\beta}_{13} + \partial_4\bar{\beta}_{14}) \\ \partial_1\bar{\beta}_{12} - \partial_3\bar{\beta}_{23} - \partial_4\bar{\beta}_{24} \\ \partial_1\bar{\beta}_{13} + \partial_2\bar{\beta}_{23} - \partial_4\bar{\beta}_{34} \\ \partial_1\bar{\beta}_{14} + \partial_3\bar{\beta}_{24} + \partial_3\bar{\beta}_{34} \end{pmatrix} + \begin{pmatrix} \bar{\kappa}_1 \\ \bar{\kappa}_2 \\ \bar{\kappa}_3 \\ \bar{\kappa}_4 \end{pmatrix} \tag{82}$$

In the above, we have used the following abbreviation: $[\bar{\beta}] = -[\beta]$. In the absence of electricity and time dependent effects, i.e. in only magnetostatic cases, we may set $\alpha = 0$ and all the terms corresponding to the time coordinate may be eliminated. In this case, we may write:

$$\begin{pmatrix} \kappa_1 \\ \kappa_2 \\ \kappa_3 \end{pmatrix} = - \begin{pmatrix} -\left(\partial_2 \bar{\beta}_{12} + \partial_3 \bar{\beta}_{13}\right) \\ \partial_1 \bar{\beta}_{12} - \partial_3 \bar{\beta}_{23} \\ \partial_1 \bar{\beta}_{13} + \partial_2 \bar{\beta}_{23} \end{pmatrix} + \begin{pmatrix} \bar{\kappa}_1 \\ \bar{\kappa}_2 \\ \bar{\kappa}_3 \end{pmatrix} \tag{83}$$

One may now write $\boldsymbol{\kappa}$, using the axial vector field $\boldsymbol{\tau}$ corresponding to the skew-symmetric tensor $\bar{\boldsymbol{\beta}}$, as:

$$\boldsymbol{\kappa} = \nabla \times \boldsymbol{\tau} + \bar{\boldsymbol{\kappa}} \tag{84}$$

where, the components of $\boldsymbol{\tau}$ are: $\tau_1 = \bar{\beta}_{23}$, $\tau_2 = -\bar{\beta}_{13}$ and $\tau_3 = \bar{\beta}_{12}$. In the following, we make use of this decomposition to model magnetoelasticity for static cases. Under the action of a magnetic field ($\mathbf{B}$) in a magnetic material, magnetic dipole moments are induced in the body, the macroscopic density of which is known as magnetic polarization or magnetization ($\mathbf{M}$). One may define a magnetic intensity field ($\mathbf{H}$) as:

$$\mathbf{H} = \frac{\mathbf{B}}{\mu_0} - \mathbf{M} \tag{85}$$

where, $\mu_0$ is the permeability of free space. The magnetic field may be obtained from the curl of a vector field known as the magnetic vector potential ($\mathbf{A}$): $\mathbf{B} = \nabla \times \mathbf{A}$. Under the action of mechanical loading, $\mathbf{H}$ depends on strain and strain gradient. Thus equation (85) must be modified with appropriate strain and strain gradient terms. Keeping this in mind, we first interpret $\boldsymbol{\tau} = -\nu \frac{\mathbf{A}}{\mu_0}$ and $\bar{\boldsymbol{\kappa}} = \nu \mathbf{M}$, where $\nu$ is a proportionality constant used to make the units consistent. In our conformal gauge theory, magnetomechanical coupling is brought about through Weyl's condition in the Lagrangian, i.e. via a coupling of $\boldsymbol{\tau}$, $\bar{\boldsymbol{\kappa}}$ and $\nabla C$. One may also interpret $\boldsymbol{\kappa} = -\mathbf{H}$, where $\mathbf{H}$ is the magnetic intensity field in the absence of mechanical loading. In magnetoelastic cases, $\mathbf{u}$, $\boldsymbol{\tau}$ and $\bar{\boldsymbol{\kappa}}$ may be solved from the Euler-Lagrange equations and using these $\mathbf{A}$ and $\mathbf{M}$ may be computed.

In summary, for electrostatic and magnetostatic conditions, the Lagrange density $L$ is a function of $\mathbf{C}$, $\lambda$, $\bar{\gamma}$, $\boldsymbol{\tau}$ and $\bar{\boldsymbol{\kappa}}$. Variation of $L$ may be written as:

$$\begin{aligned}\delta L &= \mathcal{D}_{\mathbf{C}}L : \delta\mathbf{C} + \mathcal{D}_{\lambda}L\delta\lambda + \mathcal{D}_{\bar{\gamma}}L \cdot \delta\bar{\gamma} + \mathcal{D}_{\tau}L \cdot \delta\boldsymbol{\tau} + \mathcal{D}_{\bar{\kappa}}L \cdot \delta\bar{\boldsymbol{\kappa}} \\ &= (\mathbf{S}/2) : \delta\mathbf{C} + P\delta\lambda + \mathbf{Q} \cdot \delta\bar{\gamma} + \mathbf{U} \cdot \delta\boldsymbol{\tau} + \mathbf{V} \cdot \delta\bar{\boldsymbol{\kappa}}\end{aligned} \quad (86)$$

We have made use of the following abbreviations: $\mathbf{S} := 2\mathcal{D}_{\mathbf{C}}L$, $P := \mathcal{D}_{\lambda}L$, $\mathbf{Q} := \mathcal{D}_{\bar{\gamma}}L$, $\mathbf{U} := \mathcal{D}_{\tau}L$ and $\mathbf{V} := \mathcal{D}_{\bar{\kappa}}L$.

An important point needs to be noted here. To derive electromechanical or magnetomechanical coupling through the conformal gauge theory, we have departed from the conventional ideas in electromagnetism, wherein $\bar{\boldsymbol{\kappa}}$ is interpreted as the electromagnetic 4-potential which contains the electric scalar potential and the magnetic vector potential in the 4-dimensional case. Considering such a case, if we try to impose Weyl's condition through the Lagrangian, then the electric potential and magnetic potential will be directly coupled with $\mathbf{C}$ and $\nabla\mathbf{C}$ which is clearly unphysical. This is so as we know that only electric and magnetic fields affect strain and not the magnitudes of the associated potentials.

## 9 Equations for flexomagnetism

Flexomagnetism is a magnetoelastic coupling between strain gradient and magnetization. This effect is expected to be pronounced in the nanoscale where very high strain gradients are common. Lukashev and Sabirianov (2010) have calculated flexomagnetic coefficients in $\text{Mn}$ - based antiperovskites, e.g. $\text{Mn}_3\text{GaN}$, through ab-initio simulations and the density functional theory.

To model flexomagnetism, we need to add an extra term in the Lagrange density which remains invariant under the transformation $\grave{\mathbf{C}} = e^{f(\mathbf{X})}\mathbf{C}$. As with the case for flexoelectricity, we write the gauge invariant term for flexomagnetism ($\Psi_{flm}$) in the following way:

$$\Psi_{flm} = \bar{h}_{\alpha\mu\nu pqr}\left[\partial_{\alpha}C_{\mu\nu} - \kappa_{\alpha}C_{\mu\nu}\right]\left[\partial_{p}C_{qr} - \kappa_{p}C_{qr}\right](\det\mathbf{C})^{-\frac{2}{3}} \quad (87)$$

$\overline{\mathbf{h}}$ is a sixth order constitutive tensor and $\mathbf{\kappa}$ is as prescribed in equation (84). We model flexomagnetism through a coupling between $\nabla \mathbf{C}$ and $\mathbf{\kappa}$ in $\Psi_{flm}$. Variation of $\Psi_{flm}$ may be written as:

$$\delta \Psi_{flm} = \overline{h}_{\alpha\mu\nu pqr} \left( \partial_\alpha \delta C_{\mu\nu} - \delta\kappa_\alpha C_{\mu\nu} - \kappa_\alpha \delta C_{\mu\nu} \right) \left( \partial_p C_{qr} - \kappa_p C_{qr} \right) (\det \mathbf{C})^{-\frac{2}{3}}$$
$$+ \overline{h}_{pqr\alpha\mu\nu} \left( \partial_p C_{qr} - \kappa_p C_{qr} \right) \left( \partial_\alpha \delta C_{\mu\nu} - \delta\kappa_\alpha C_{\mu\nu} - \kappa_\alpha \delta C_{\mu\nu} \right) (\det \mathbf{C})^{-\frac{2}{3}} \quad (88)$$
$$- \frac{2}{3} \overline{h}_{\alpha mnpqr} C_{\mu\nu}^{-1} \delta C_{\mu\nu} \left( \partial_\alpha C_{mn} - \kappa_\alpha C_{mn} \right) \left( \partial_p C_{qr} - \kappa_p C_{qr} \right) (\det \mathbf{C})^{-\frac{2}{3}}$$

Using this, we add extra terms $\mathbf{S}_{add,flm}$, $\mathbf{U}_{add,flm}$ and $\mathbf{V}_{add,flm}$ with $\mathbf{S}$, $\mathbf{U}$ and $\mathbf{V}$ respectively, in the presence of the flexomagnetic effect (see equation 86). We may express $\mathbf{S}_{add,flm}$ as follows:

$$\left( S_{add,flm} \right)_{\mu\nu} = -2\partial_\alpha \left[ \left( \overline{h}_{\alpha\mu\nu pqr} + \overline{h}_{pqr\alpha\mu\nu} \right) \left( \partial_p C_{qr} - \kappa_p C_{qr} \right) (\det \mathbf{C})^{-\frac{2}{3}} \right]$$
$$- 2 \left( \overline{h}_{\alpha\mu\nu pqr} + \overline{h}_{pqr\alpha\mu\nu} \right) \kappa_\alpha \left( \partial_p C_{qr} - \kappa_p C_{qr} \right) (\det \mathbf{C})^{-\frac{2}{3}} \quad (89)$$
$$- \frac{4}{3} \overline{h}_{\alpha mnpqr} C_{\mu\nu}^{-1} \left( \partial_\alpha C_{mn} - \kappa_\alpha C_{mn} \right) \left( \partial_p C_{qr} - \kappa_p C_{qr} \right) (\det \mathbf{C})^{-\frac{2}{3}}$$

$\mathbf{U}_{add,flm}$ is given as:

$$\left( U_{add,flm} \right)_n = \partial_m \left[ \varepsilon_{\alpha mn} \left( h_{\alpha\mu\nu pqr} + h_{pqr\alpha\mu\nu} \right) C_{\mu\nu} \left( \partial_p C_{qr} - \kappa_p C_{qr} \right) (\det \mathbf{C})^{-\frac{2}{3}} \right] \quad (90)$$

Finally $\mathbf{V}_{add,flm}$ is given as:

$$\left( V_{add,flm} \right)_\alpha = -\left( \overline{h}_{\alpha\mu\nu pqr} + \overline{h}_{pqr\alpha\mu\nu} \right) C_{\mu\nu} \left( \partial_p C_{qr} - \kappa_p C_{qr} \right) (\det \mathbf{C})^{-\frac{2}{3}} \quad (91)$$

**10 Extension to piezomagnetism**

Piezomagnetism is a magnetomechanical coupling seen in some antiferromagnetic materials. It is a linear coupling of strain and magnetization. Dzialoshinskii (1957) has predicted the possibility of piezomagnetic effects even before an experimental confirmation. Moriya (1959) has theoretically shown that piezomagnetic effects are present in $MnF_2$, $FeF_2$ and $CoF_2$ with the highest magnitude being in $CoF_2$. Borovik-Romanov (1960) have experimentally confirmed the existence of piezomagnetic effects in $CoF_2$ and $MnF_2$ upon the application of shear stresses and also measured piezomagnetic coefficients for these materials.

Piezomagnetism may be modeled through the conformal gauge theory based on a contraction of $(\nabla \mathbf{C} - \boldsymbol{\kappa} \mathbf{C})(\det \mathbf{C})^{-\frac{1}{3}}$ with a third order constitutive tensor, which is invariant under $`\mathbf{C} = e^{f(\mathbf{X})} \mathbf{C}$. The additional gauge invariant term in the Lagrange density may be written in the following way:

$$\Psi_{pm} = \overline{e}_{\alpha\mu\nu} \left( \partial_\alpha C_{\mu\nu} - \kappa_\alpha C_{\mu\nu} \right) (\det \mathbf{C})^{-\frac{1}{3}} \tag{92}$$

with $\overline{\mathbf{e}}$ being the third order constitutive tensor. In cases where strain gradient effects are small, $\Psi_{pm}$ may be expressed approximately as: $\Psi_{pm} \simeq -\overline{e}_{\alpha\mu\nu} \kappa_\alpha C_{\mu\nu} (\det \mathbf{C})^{-\frac{1}{3}}$ which shows a close correspondence with the piezomagnetic energy density ($\Psi_{classical,pm}$) for the small deformation case: $\Psi_{classical,pm} = \tilde{\tilde{e}}_{\alpha\mu\nu} M_\alpha \varepsilon_{\mu\nu}$, where $\tilde{\tilde{\mathbf{e}}}$ is a constant third order tensor. We may now write the variation of $\Psi_{pm}$ as:

$$\begin{aligned}\delta\Psi_{pm} &= \overline{e}_{\alpha\mu\nu} \left( \partial_\alpha \delta C_{\mu\nu} - \delta\kappa_\alpha C_{\mu\nu} - \kappa_\alpha \delta C_{\mu\nu} \right) (\det \mathbf{C})^{-\frac{1}{3}} \\ &- \frac{1}{3} (\det \mathbf{C})^{-\frac{1}{3}} C^{-1}_{\mu\nu} \delta C_{\mu\nu} \overline{e}_{pqr} \left( \partial_p C_{qr} - \kappa_p C_{qr} \right)\end{aligned} \tag{93}$$

In order to write piezomagnetic equations of equilibrium, the extra terms $\mathbf{S}_{add,pm}$, $\mathbf{U}_{add,pm}$ and $\mathbf{V}_{add,pm}$ must be suitably added with $\mathbf{S}$, $\mathbf{U}$ and $\mathbf{V}$ respectively (see equation 86). $\mathbf{S}_{add,pm}$ can be written as:

$$\left( S_{add,pm} \right)_{\mu\nu} = -2\partial_\alpha \left[ \overline{e}_{\alpha\mu\nu} (\det \mathbf{C})^{-\frac{1}{3}} \right] - 2(\det \mathbf{C})^{-\frac{1}{3}} \left[ \overline{e}_{\alpha\mu\nu} \kappa_\alpha + \frac{1}{3} C^{-1}_{\mu\nu} \overline{e}_{pqr} \left( \partial_p C_{qr} - \kappa_p C_{qr} \right) \right] \tag{94}$$

$\mathbf{U}_{add,pm}$ takes the form:

$$\left(U_{add,\text{pm}}\right)_n = \partial_m\left[\varepsilon_{\alpha mn}e_{\alpha\mu\nu}C_{\mu\nu}\left(\det\mathbf{C}\right)^{-\frac{1}{3}}\right] \tag{95}$$

$\mathbf{V}_{add,\text{pm}}$ can be written in the following manner.

$$\left(V_{add,\text{pm}}\right)_\alpha = -\overline{e}_{\alpha\mu\nu}C_{\mu\nu}\left(\det\mathbf{C}\right)^{-\frac{1}{3}} \tag{96}$$

## 11 Modelling magnetostriction

Discovered by James Joule in 1842, magnetostriction is a magnetomechanical coupling phenomenon seen in ferromagnetic materials. Here, strain is coupled with the quadratic of magnetization. These materials are used as sensors and actuators as they are able to transform magnetic energy into mechanical energy and vice-versa. A few materials which show high magnetostriction effects are: Terfenol-D, Alfenol, Galfenol, Metglas, Cobalt ferrite etc. Furthmüller *et al*. (1987) have developed a theory of magnetostriction for amorphous and polycrystalline ferromagnets. Sandlund *et al*. (1994) have experimentally shown very large magnetostriction effects for composites of Terfenol-D alloys, useful in high-frequency applications. A model for magnetostriction through a minimization of energy is furnished by DeSimone and James (1997) and James and Kinderlehrer (1994). Hall (1959) experimentally measured magnetostriction coefficients for many alloys, e.g. NiFe, SiFe, AlFe, CoNi and CoFe.

Similar to the case with electrostriction, one may attempt to model magnetostriction using two terms $C_{pq}^{-1}\left(\partial_k C_{pq} - \kappa_k C_{pq}\right)$ and $\left(\partial_l C_{ij} - \kappa_l C_{ij}\right)\left(\det\mathbf{C}\right)^{-\frac{1}{3}}$ which remain invariant under the transformation $`\mathbf{C} = e^{f(\mathbf{X})}\mathbf{C}$. We may write the extra gauge invariant term in the Lagrange density as:

$$\Psi_{ms} = \overline{q}_{ijkl}\left(C_{pq}^{-1}\partial_k C_{pq} - 3\kappa_k\right)\left(\partial_l C_{ij} - \kappa_l C_{ij}\right)\left(\det\mathbf{C}\right)^{-\frac{1}{3}} \tag{97}$$

where, $\bar{\mathbf{q}}$ is a fourth order constitutive tensor. When strain gradient effects are small, $\Psi_{ms}$ takes the following form: $\Psi_{ms} = 3\bar{q}_{ijkl}\kappa_k \kappa_l C_{ij} (\det \mathbf{C})^{-\frac{1}{3}}$, which shows a close resemblance with the standard energy density of magnetostriction given by: $\Psi_{classical,ms} = \tilde{q}_{ijkl} M_k M_l \varepsilon_{ij}$, where, $\tilde{\mathbf{q}}$ is a constant fourth order tensor. Variation of $\Psi_{ms}$ is written as:

$$\delta\Psi_{ms} = \bar{q}_{ijal}\left(-C^{-1}_{p\mu}C^{-1}_{qv}\partial_\alpha C_{pq}\delta C_{\mu v} + C^{-1}_{\mu v}\partial_\alpha \delta C_{\mu v} - 3\delta\kappa_\alpha\right)\left(\partial_l C_{ij} - \kappa_l C_{ij}\right)(\det \mathbf{C})^{-\frac{1}{3}}$$
$$+ \bar{q}_{\mu v l\alpha}\left(C^{-1}_{pq}\partial_l C_{pq} - 3\kappa_l\right)\left(\partial_\alpha \delta C_{\mu v} - \delta\kappa_\alpha C_{\mu v} - \kappa_\alpha \delta C_{\mu v}\right)(\det \mathbf{C})^{-\frac{1}{3}} \quad (98)$$
$$-\frac{1}{3}\bar{q}_{ijkl}C^{-1}_{\mu v}\delta C_{\mu v}\left(C^{-1}_{pq}\partial_k C_{pq} - 3\kappa_k\right)\left(\partial_l C_{ij} - \kappa_l C_{ij}\right)(\det \mathbf{C})^{-\frac{1}{3}}$$

We need to add the additional terms $\mathbf{S}_{add,ms}$, $\mathbf{U}_{add,ms}$ and $\mathbf{V}_{add,ms}$ with $\mathbf{S}$, $\mathbf{U}$ and $\mathbf{V}$ as in equation (86) to account for the magnetostrictive effect. We may write $\mathbf{S}_{add,ms}$ as:

$$\left(S_{add,ms}\right)_{\mu v} = -2\bar{q}_{ijal}C^{-1}_{p\mu}C^{-1}_{qv}\partial_\alpha C_{pq}\left(\partial_l C_{ij} - \kappa_l C_{ij}\right)(\det \mathbf{C})^{-\frac{1}{3}}$$
$$- 2\bar{q}_{\mu v l\alpha}\left(C^{-1}_{pq}\partial_l C_{pq} - 3\kappa_l\right)\kappa_\alpha (\det \mathbf{C})^{-\frac{1}{3}}$$
$$- \frac{2}{3}\bar{q}_{ijkl}C^{-1}_{\mu v}\left(C^{-1}_{pq}\partial_k C_{pq} - 3\kappa_k\right)\left(\partial_l C_{ij} - \kappa_l C_{ij}\right)(\det \mathbf{C})^{-\frac{1}{3}} \quad (99)$$
$$- 2\partial_\alpha \left[\bar{q}_{ijal}C^{-1}_{\mu v}\left(\partial_l C_{ij} - \kappa_l C_{ij}\right)(\det \mathbf{C})^{-\frac{1}{3}} + \bar{q}_{\mu v l\alpha}\left(C^{-1}_{pq}\partial_l C_{pq} - 3\kappa_l\right)(\det \mathbf{C})^{-\frac{1}{3}}\right]$$

$\mathbf{U}_{add,ms}$ is expressible as:

$$\left(U_{add,es}\right)_n = \partial_m \left[3\varepsilon_{amn}q_{ijal}\left(\partial_l C_{ij} - \gamma_l C_{ij}\right)(\det \mathbf{C})^{-\frac{1}{3}} + \varepsilon_{amn}q_{\mu v l\alpha}\left(C^{-1}_{pq}\partial_l C_{pq} - 3\gamma_l\right)C_{\mu v}(\det \mathbf{C})^{-\frac{1}{3}}\right] \quad (100)$$

Finally, the expression for $\mathbf{V}_{add,ms}$ is:

$$\left(V_{add,ms}\right)_\alpha = -\left[3\bar{q}_{ijal}\left(\partial_l C_{ij} - \kappa_l C_{ij}\right) + \bar{q}_{\mu v l\alpha}\left(C^{-1}_{pq}\partial_l C_{pq} - 3\kappa_l\right)C_{\mu v}\right](\det \mathbf{C})^{-\frac{1}{3}} \quad (101)$$

## 12. An illustrative example for piezoelectricity

We now try and gain some insight into how the proposed piezoelectricity governing equations work. First, the material parameters of our theory are determined from a comparison of the linearized version of the proposed governing equations with those of the classical (linear) piezoelectricity theory. We then consider a specific type of isochoric deformation and analytically solve for the electric field using our nonlinear governing equations, the linearized version and the classical piezoelectricity equations. Based on the solutions so obtained, we try to understand how our proposal works vis-à-vis the classical piezoelectricity.

### 12.1 Linearized governing equations

Since piezoelectricity occurs in non-centrosymmetric materials only, we choose the following strain energy density function (also see Section 6):

$$\Psi = \frac{1}{2} E_{\mu\nu} \mathbb{C}_{\mu\nu\alpha\beta} E_{\alpha\beta} + e_{\alpha\mu\nu} \left( \partial_\alpha C_{\mu\nu} - \gamma_\alpha C_{\mu\nu} \right) (\det \mathbf{C})^{-\frac{1}{3}}$$
$$+ K_{\alpha\beta} \left( C_{pq}^{-1} \partial_\alpha C_{pq} - 3\gamma_\alpha \right) \left( C_{rs}^{-1} \partial_\beta C_{rs} - 3\gamma_\beta \right) \tag{102}$$
$$+ \frac{c_4}{4} \left( \partial_k \overline{\gamma}_i - \partial_i \overline{\gamma}_k \right) \left( \partial_k \overline{\gamma}_i - \partial_i \overline{\gamma}_k \right)$$

Here, $\mathbb{C}$ and $\mathbf{K}$ are fourth and second order constant tensors respectively and $\mathbf{E} = \frac{1}{2}(\mathbf{C} - \mathbf{I})$ the Green-Lagrange strain tensor. The first term in equation (102) is a simple extension of generalized Hooke's law for finite deformation. The third term in equation (102) is an additional gauge invariant term considered in the strain energy density. We will show that, with this strain energy density, the linearized governing equations (without considering the anti-exact part of the 1-form field) will be exactly the classical piezoelectricity equations under isochoric deformation conditions. The Euler-Lagrange equations derived from Hamilton's principle (equation 39) with $\mathcal{L} = -\Psi$ can be written as:

$$\partial_j (\mathbf{FS})_{ij} + f_\alpha = 0 \tag{103}$$

$$\partial_\alpha \left[ e_{\alpha\mu\nu} C_{\mu\nu} (\det \mathbf{C})^{-\frac{1}{3}} + 3\left(K_{\alpha\beta} + K_{\beta\alpha}\right)\left(C_{rs}^{-1} \partial_\beta C_{rs} - 3\gamma_\beta\right) \right] = 0 \qquad (104)$$

$$c_4 \partial_\beta \left(\partial_\beta \bar{\gamma}_\alpha - \partial_\alpha \bar{\gamma}_\beta\right) + e_{\alpha\mu\nu} C_{\mu\nu} (\det \mathbf{C})^{-\frac{1}{3}} + 3\left(K_{\alpha\beta} + K_{\beta\alpha}\right)\left(C_{rs}^{-1} \partial_\beta C_{rs} - 3\gamma_\beta\right) = 0 \qquad (105)$$

where, the second Piola-Kirchoff stress $\mathbf{S}$ is given by:

$$\begin{aligned}
S_{\mu\nu} &= \mathbb{C}_{\mu\nu\alpha\beta} E_{\alpha\beta} - 2\partial_\alpha \left[e_{\alpha\mu\nu}(\det \mathbf{C})^{-\frac{1}{3}}\right] - 2(\det \mathbf{C})^{-\frac{1}{3}} \left[e_{\alpha\mu\nu}\gamma_\alpha + \frac{1}{3} C_{\mu\nu}^{-1} e_{pqr}\left(\partial_p C_{qr} - \gamma_p C_{qr}\right)\right] \\
&\quad - 2\left(K_{\alpha\beta} + K_{\beta\alpha}\right) C_{p\mu}^{-1} C_{q\nu}^{-1} \partial_\alpha C_{pq} \left(C_{rs}^{-1} \partial_\beta C_{rs} - 3\gamma_\beta\right) \\
&\quad - 4\partial_\alpha \left[\left(K_{\alpha\beta} + K_{\beta\alpha}\right) C_{\mu\nu}^{-1} \left(C_{rs}^{-1} \partial_\beta C_{rs} - 3\gamma_\beta\right)\right]
\end{aligned} \qquad (106)$$

Now, using the approximations $\mathbf{C} \approx \mathbf{I} + 2\boldsymbol{\varepsilon}$, $\mathbf{C}^{-1} \approx \mathbf{I} - 2\boldsymbol{\varepsilon}$, $\det \mathbf{C} \approx 1 + 2 tr\boldsymbol{\varepsilon}$, and neglecting $\nabla \mathbf{C}$ and all nonlinear terms from equation (106), one may write the linearized stress $\mathbf{S}$ as follows:

$$S_{\mu\nu} \approx \mathbb{C}_{\mu\nu\alpha\beta} \varepsilon_{\alpha\beta} - 2\left(e_{\alpha\mu\nu} - \frac{1}{3} e_{\alpha\kappa\kappa} \delta_{\mu\nu}\right) \gamma_\alpha \qquad (107)$$

Similarly, the linearized versions of equations (104) and (105) are presented below.

$$\partial_\alpha \left[2 e_{\alpha\mu\nu} \varepsilon_{\mu\nu}^{dev} + 9\left(K_{\alpha\beta} + K_{\beta\alpha}\right) \gamma_\beta\right] = 0 \qquad (108)$$

$$c_4 \partial_\beta \left(\partial_\beta \bar{\gamma}_\alpha - \partial_\alpha \bar{\gamma}_\beta\right) + e_{\alpha\mu\nu}\left(\delta_{\mu\nu} + 2\varepsilon_{\mu\nu}^{dev}\right) - 9\left(K_{\alpha\beta} + K_{\beta\alpha}\right) \gamma_\beta = 0 \qquad (109)$$

## 12.2 Determination of material parameters

We express the material parameters of our theory in terms of those in classical piezoelectricity using an equivalence of the governing equations in the linearized setup and under isochoric deformation condition. The Lagrange density for classical piezoelectricity may be written as (see Tiersten, 2013):

$$\mathcal{L} = -\left(\frac{1}{2} \tilde{\mathbb{C}}_{ijkl} \varepsilon_{ij} \varepsilon_{kl} - \tilde{e}_{ijk} E_i \varepsilon_{jk} - \frac{1}{2} \tilde{K}_{ij} E_i E_j\right) \qquad (110)$$

where, $\tilde{\mathbb{C}}$, $\tilde{\mathbf{e}}$ and $\tilde{\mathbf{K}}$ denote the stiffness tensor, the piezoelectric constant tensor and the dielectric constant tensor respectively. Considering isochoric deformation and using $\tilde{\mathbb{C}}_{ijkl} = \tilde{\mathbb{C}}_{klij}$ and $\tilde{K}_{ij} = \tilde{K}_{ji}$, we may write the variation of $\mathcal{L}$ as follows:

$$\delta\mathcal{L} = -\left[\left(\tilde{\mathbb{C}}_{ijkl} - \frac{1}{3}\tilde{\mathbb{C}}_{mmkl}I_{ij}\right)\varepsilon_{kl}^{dev} + \left(\tilde{e}_{kij} - \frac{1}{3}\tilde{e}_{kpp}I_{ij}\right)E_k\right]\delta\varepsilon_{ij} - \tilde{K}_{ij}E_j\delta E_i \tag{111}$$

The Euler-Lagrange equations derived from Hamilton's principle (equation 39) may be given by:

$$\nabla \cdot \tilde{\boldsymbol{\sigma}} + \mathbf{f} = \mathbf{0} \tag{112}$$

$$\nabla \cdot \tilde{\mathbf{D}} = 0 \tag{113}$$

where,

$$\tilde{\sigma}_{ij} = \tilde{\mathbb{C}}_{ijkl}\varepsilon_{kl}^{dev} - \left(\tilde{e}_{mij} - \frac{1}{3}\tilde{e}_{mkk}\delta_{ij}\right)E_m \tag{114}$$

$$\tilde{D}_i = \tilde{e}_{ijk}\varepsilon_{jk}^{dev} + \tilde{K}_{ij}E_j \tag{115}$$

Similarly, one may derive the Euler-Lagrange equations (using equation 39) from the strain energy density given in equation (102) without considering the anti-exact part of the 1-form field. Upon linearization, the Euler-Lagrange equations take the following form:

$$\partial_\nu \sigma_{\mu\nu} + \rho b_\nu = 0 \tag{116}$$

$$\partial_\alpha\left[2e_{\alpha\mu\nu}\varepsilon_{\mu\nu}^{dev} + 9\left(K_{\alpha\beta} + K_{\beta\alpha}\right)\partial_\beta\lambda\right] = 0 \tag{117}$$

where,

$$\sigma_{\mu\nu} = \mathbb{C}_{\mu\nu\alpha\beta}\varepsilon_{\alpha\beta}^{dev} - 2\left(e_{\alpha\mu\nu} - \frac{1}{3}e_{\alpha\kappa\kappa}\delta_{\mu\nu}\right)\partial_\alpha\lambda \tag{118}$$

Note that, if we choose $\mathbb{C} = \tilde{\mathbb{C}}$, $\mathbf{e} = -\tilde{\mathbf{e}}/(2\varepsilon_0\mu)$, $\lambda = \varepsilon_0\mu\varphi$ (see Section 4), $\mathbf{K}$ symmetric and $\mathbf{K} = -\tilde{\mathbf{K}}/(9\varepsilon_0\mu)$, then the linearized equations of the proposed theory under isochoric deformation (equations 116 and 117) are identical with the classical ones (equations 112 and 113).

## 12.3 Solution for isochoric deformation

Consider the deformation map:

$$x_1 = X_1 + p(X_2), \quad x_2 = X_2, \quad x_3 = X_3 \tag{119}$$

$X_1$, $X_2$ and $X_3$ are the coordinates assigned to the reference configuration. The right Cauchy-Green matrix is given below.

$$\mathbf{C} = \begin{bmatrix} 1 & p' & 0 \\ p' & 1+p'^2 & 0 \\ 0 & 0 & 1 \end{bmatrix} \tag{120}$$

Here, $p' = \dfrac{dp}{dX_2}$. This corresponds to isochoric deformation as $\det(\mathbf{C}) = 1$. The material chosen here is PZT-5H which is transversely isotropic and exhibits piezoelectric properties. The material constants are presented in Table 2 (see Zhu *et al.*, 1998). Note that $X_1$, $X_2$ and $X_3$ are chosen along the material axes with respect to which the material parameters are referred.

Table 2: Material parameters for PZT-5H

| $\tilde{\mathbb{C}}_{1111}$ (GPa) | $\tilde{\mathbb{C}}_{1122}$ (GPa) | $\tilde{\mathbb{C}}_{1133}$ (GPa) | $\tilde{\mathbb{C}}_{3333}$ (GPa) | $\tilde{\mathbb{C}}_{2323}$ (GPa) |
|---|---|---|---|---|
| 113 | 67.3 | 66.9 | 98.9 | 20.1 |

| $\tilde{e}_{113}$ $(C/m^2)$ | $\tilde{e}_{311}$ $(C/m^2)$ | $\tilde{e}_{333}$ $(C/m^2)$ | $\tilde{K}_{11}$ | $\tilde{K}_{33}$ |
|---|---|---|---|---|
| 14.8 | −5.26 | 23.5 | 1550 | 1691 |

$\mathbb{C}$ has major and minor symmetries (i.e. $\mathbb{C}_{ijkl} = \mathbb{C}_{klij}$, $\mathbb{C}_{ijkl} = \mathbb{C}_{ijlk} = \mathbb{C}_{jikl}$) and the relations $\mathbb{C}_{1212} = (\mathbb{C}_{1111} - \mathbb{C}_{1122})/2$, $\tilde{e}_{223} = \tilde{e}_{113}$, $\tilde{e}_{322} = \tilde{e}_{311}$ hold. We determine the material parameters of our theory using the correspondence equations furnished in Section 12.2.

### 12.3.1 Insights from nonlinear equations

If we choose $\gamma_1 = 0$, $\gamma_2 = 0$ and $\gamma_3 = g(X_2)$, then for the deformation prescribed in equation (119) and material properties presented in Table 2, equation (104) is satisfied. Noting the identity $\partial_\beta \overline{\gamma}_\alpha - \partial_\alpha \overline{\gamma}_\beta = \partial_\beta \gamma_\alpha - \partial_\alpha \gamma_\beta$, equation (105) is recast as:

$$c_4 \partial_\beta \left( \partial_\beta \gamma_\alpha - \partial_\alpha \gamma_\beta \right) + e_{\alpha\mu\nu} C_{\mu\nu} \left( \det \mathbf{C} \right)^{-\frac{1}{3}} + 6 K_{\alpha\beta} \left( C_{rs}^{-1} \partial_\beta C_{rs} - 3 \gamma_\beta \right) = 0 \qquad (121)$$

It may be verified that for $\alpha = 1$ and $\alpha = 2$, equation (121) is satisfied and for $\alpha = 3$, equation (121) reduces to the following second order ordinary differential equation (ODE):

$$g'' - \frac{18 K_{33}}{c_4} g = -\frac{1}{c_4} \left( e_{311} + e_{322} + e_{333} + e_{322} p'^2 \right) \qquad (122)$$

Here, $g'' = \dfrac{d^2 g}{dX_2^2}$. Considering $p(X_2) = \sin(n\pi X_2 / L)$ and using the relations $\mathbf{e} = -\tilde{\mathbf{e}}/(2\varepsilon_0 \mu)$ and $K_{33} = -\tilde{K}_{33}/(9\varepsilon_0 \mu)$ (see Section 12.2), equation (122) may be recast as:

$$g'' + k_1^2 g = k_2 + k_3 \cos(k_4 x) \qquad (123)$$

where, $k_1 = \sqrt{2\tilde{K}_{33}/(c_4 \varepsilon_0 \mu)}$, $k_2 = \left( \tilde{e}_{311} + \tilde{e}_{322} + \tilde{e}_{333} \right)/(2\varepsilon_0 \mu c_4) + \tilde{e}_{322} n^2 \pi^2 /(4 L^2 \varepsilon_0 \mu c_4)$, $k_3 = \tilde{e}_{322} n^2 \pi^2 /(4 L^2 \varepsilon_0 \mu c_4)$ and $k_4 = 2n\pi/L$. Equation (123) is solved by direct and inverse Laplace transforms and the solution may be expressed as:

$$g = g_0 \cos(k_1 x) + \frac{g_0'}{k_1} \sin k_1 x + \frac{k_2}{k_1^2} (1 - \cos k_1 x) + \frac{k_3}{(k_4^2 - k_1^2)} (\cos k_1 x - \cos k_4 x) \qquad (124)$$

where $g_0 = g(0)$ and $g'_0 = g'(0)$. Choosing $g_0 = g'_0 = 0$, we now plot $g$ for various $n$. Apart from parameters specified in Table 2, the additional parameters considered here are as follows: $\varepsilon_0 \mu = 1 \text{V}^{-1}$, $c_4 = 1 \text{Nm}^2$, $L = 1 m$.

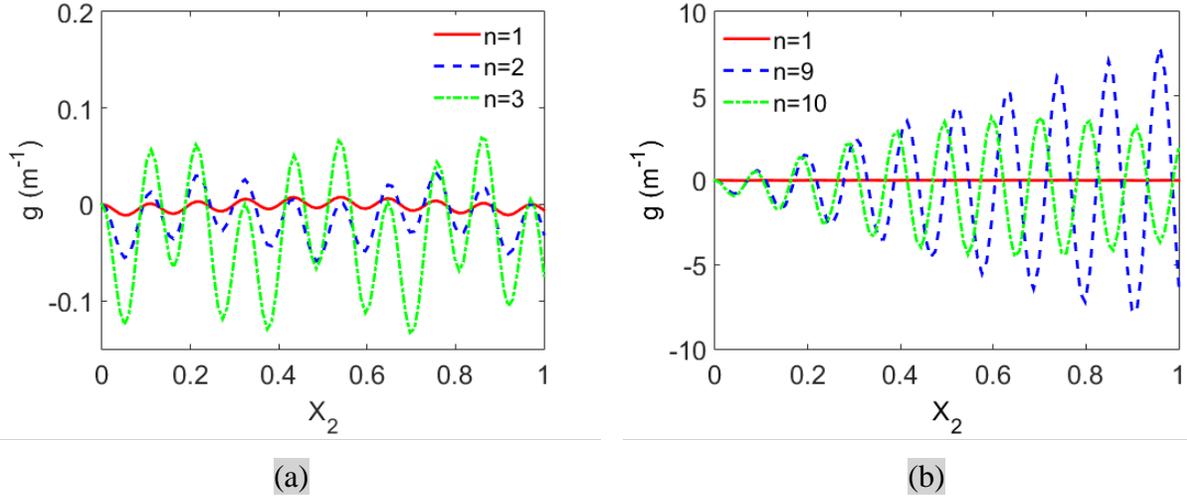

(a)            (b)

Figure 1: Plot of $g$ with $X_2$ for various $n$ (a) for $n = 1, 2$ and $3$ (b) for $n = 1, 9$ and $10$

From equation (124), one notes that as $k_4$ approaches $k_1$, the solution blows off. We may find a critical value of $n$ ($n = n_{crit}$) corresponding to such a singularity by setting $k_4 = k_1$:

$$n_{crit} = \frac{L}{\sqrt{2\pi}} \sqrt{\frac{\tilde{K}_{33}}{c_4 \varepsilon_0 \mu}} = 9.26 \approx 9 \tag{125}$$

As $n$ is an integer, we may only choose the closest integer instead of the calculated numerical value in the above equation. We may also note that, for very large $n$, the third term of equation (124) dominates and the value of $g$ becomes very large.

Figure (1a) shows that, as $n$ increases, the absolute maximum and absolute minimum values of $g$ increase. However, for $n = 9$ (see equation 125), high values of $g$ are seen (Figure 1b) which pertain to near-resonance as the applied spatial frequency is close to one of the natural frequencies of the system. This indicates that, one may generate more electric displacement by an appropriate choice of the deformation field.

### 12.3.2 Linearized case

The non-zero terms in the linearized strain matrix are $\varepsilon_{12} = \varepsilon_{21} = p'/2$. With this, equation (117) takes the following form:

$$\partial_\alpha \left[ 9 \left( K_{\alpha\beta} + K_{\beta\alpha} \right) \partial_\beta \lambda \right] = 0 \tag{126}$$

As there is no driving from mechanical deformation for the evolution of $\partial_\beta \lambda$, the natural choice for the solution will be $\partial_\beta \lambda = 0$ which satisfies equation (126). This implies that if the electric field $(-\varepsilon_0 \mu \partial_\beta \lambda)$ is zero initially, it will remain zero.

### 12.3.3 Classical piezoelectricity solution

Substituting $\tilde{\mathbf{D}}$ from equation (115) to equation (113), we obtain:

$$\partial_i \left( e_{ijk} \varepsilon_{jk}^{dev} + k_{ij} E_j \right) = 0 \tag{127}$$

Note that, for material properties given in Table 2 and non-zero strain components $\varepsilon_{12} = \varepsilon_{21} = p'/2$, the driving term for the electric field is:

$$\partial_i \left( e_{ijk} \varepsilon_{jk}^{dev} \right) = 0 \tag{128}$$

Therefore, if we choose $\mathbf{E} = \mathbf{0}$ similar to the case in Section 12.3.2, equation (127) is satisfied.

### 12.3.4 Discussions

While both classical piezoelectricity and the linearized version of our theory predict that the electric field does not evolve for the deformation map given by equations (119), the nonlinear governing equations (equations 103 through 105) of our proposal result in a non-trivial evolution of the 1-form field. This shows the importance of the nonlinear terms in piezoelectricity which a linear theory of piezoelectricity does not consider, failing thereby to predict the correct evolution of electric fields in some cases. We also observe that the magnitude of $\gamma_3$ may be made larger through a proper choice of the wavenumber of the spatial deformation field (see Figure 1). This is suggestive of a procedure to enhance the energy harvesting capabilities by exploiting the nonlinear regime of piezoelectric material behavior. Of certain relevance is an experimental procedure discussed in Park *et al*. (2017) on acoustic forced vibration on PVDF thin films (a piezoelectric polymer which deforms approximately in an isochoric manner) under tension. They experimentally predict that there exists a resonance frequency for which the magnitude of generated electric field increases drastically. Such a resonance frequency for the standing wave naturally depends on the material properties of the film and the amount of applied tension. Based on the analysis presented in Section 12.3.1, we have already seen that, for some critical value of the wavenumber, the electric displacement is enhanced manifold. Therefore, the framework we have laid out should be useful in a possible explanation and predictive exploitation of such high electric fields - essentially a manifestation of nonlinear piezoelectricity. The experiment conducted in Park *et al*. (2017) will serve as a benchmark for a future validation of our theory (when we equip it with dynamics).

Yet another observation is that $\boldsymbol{\gamma}$ has a direct coupling only with terms that are isochoric. Although this is a reflection of how the theory is formulated (viz. only the local conformal invariance of the isochoric part of Lagrange density has been considered), the implication might be that contributions of the isochoric part are more pronounced vis-à-vis the dilatational part. Under pure dilatation, one may indeed imagine a deformation field that does not cause the centers of the positive and negative charges displace relative to each other. We also note that the classical piezoelectricity theory does not account for polarization as an independent field variable, even as the present approach considers both the exact and antiexact parts of the 1-form field as independent.

## 13 Concluding remarks

We have developed a conformal gauge theory of solids. Interpreting the right Cauchy-Green tensor as a metric, we have shown, upon a scaling of the metric by a constant factor, that the isochoric part of the Lagrangian remains invariant even as the volumetric part changes.

However, if the scale factor depends on position, then the isochoric part too loses its invariance. In order to restore the invariance of the isochoric part of the Lagrangian under position dependent conformal transformations, we introduce the notion of a gauge covariant derivative through a 1-form field and thus obtain a minimally replaced Lagrangian. A close similarity exists between our minimal replacement construct and the Weyl condition in non-Riemannian differential geometry. By constructing suitable energy-like terms, we have effectively imposed the Weyl condition through the Lagrangian itself. We have further augmented the Lagrangian with a term involving the exterior derivative of the 1-form field so it may evolve non-trivially. Using Hamilton's principle, we have derived the Euler-Lagrange equations and reflected at length on the nature of the physical phenomena described by these equations. Noting, for instance, the coupling between the 1-form field and the covariant derivative of the metric, the Euler-Lagrange equations derivable from a subclass of the Lagrangian bear similarity with known flexoelectricity equations if we identify the exact part of the 1-form field with the electric field and the anti-exact part with the polarization vector. Other subclasses of the Lagrangian are obtained by variously contracting the difference term from the Weyl condition. We have thus shown how to model piezoelectricity and electrostriction – all in a finite deformation setting. For magnetostatic cases, we have started with a 4-dimensional setup incorporating time and decomposed the 1-form field using the Hodge decomposition theorem. This decomposition, in the special case of magnetostatics, boils down to a curl of a pseudo-vector field and a vector field. Identifying the pseudo-vector field as magnetic potential and another part of the 1-form field as magnetization, we have shown a way to model flexomagnetism, piezomagnetism and magnetostriction. Through an example on piezoelectricity, we have illustrated how the nonlinear terms in the governing equations might affect the response.

Starting with the elasticity theory and only requiring a conformal symmetry on the action functional, our proposal may provide for a unified framework to explain many electromechanical and magnetomechanical phenomena. Nontrivial extensions of this work might be possible with the incorporation of time as the fourth coordinate and imposition of additional symmetry requirements. Among myriad other possibilities, these might explicate on the role of inertial forces on the electro-magneto-mechanical response of solids.